\documentclass[12pt]{article}
\DeclareMathAlphabet{\scr}{U}{rsfs}{m}{n}
\usepackage{latexsym,url}
\usepackage[mathscr]{eucal}
\usepackage{amsfonts}
\usepackage{amscd}
\usepackage{cite}         
\usepackage{amssymb}
\usepackage[centertags]{amsmath}
\usepackage{enumerate}
\usepackage{graphicx}
\usepackage{booktabs}
\usepackage{hyperref}
\usepackage{ulem}

\newcommand{\cleqn}{\setcounter{equation}{0}}
\newcommand{\newc}{\newcommand}
\newc{\be}{\begin{equation}}
\newc{\ee}{\end{equation}}
\newc{\bea}{\begin{eqnarray}}
\newc{\eea}{\end{eqnarray}}
\newc{\ben}{\begin{equation*}}
\newc{\een}{\end{equation*}}
\newc{\bean}{\begin{eqnarray*}}
\newc{\eean}{\end{eqnarray*}}
\newc{\ol}{\overline}
\newc{\wt}{\widetilde}
\newc{\bs}{\boldsymbol}
\newc{\m}{\mathcal}
\newc{\la}{\lambda}
\newc{\lra}{\longrightarrow}
\newc{\vp}{\varphi}
\newc{\ti}{\tilde}
\newc{\VEV}[1]{\langle#1 \rangle}

\begin{document}

\title{\hfill ~\\[-30mm]
          \hfill\mbox{\small  QFET-2017-07}\\[-3.5mm]
          \hfill\mbox{\small  SI-HEP-2017-08}\\[13mm]
       \textbf{Spontaneous CP violation in 
       multi-Higgs potentials with triplets of $\Delta(3n^2)$ and $\Delta(6n^2)$
       \\[4mm]
}}

\author{
Ivo de Medeiros Varzielas$\,^{a,b}\,$\footnote{E-mail: {\tt ivo.de@udo.edu}}
,~~
Stephen F. King$\,^a\,$\footnote{E-mail: {\tt king@soton.ac.uk}}
,\\[2mm]
Christoph Luhn$\,^{c}\,$\footnote{E-mail: {\tt christoph.luhn@uni-siegen.de}}
,~~
Thomas Neder$\,^{a,d}\,$\footnote{E-mail: {\tt neder@ific.uv.es}}\\[8mm]
$^a$\,\it{\small School of Physics and Astronomy, University of Southampton,}\\
\it{\small SO17 1BJ Southampton, United Kingdom}\\
$^b$\,\it{\small CFTP, Departamento de F\'{\i}sica, Instituto Superior T\'{e}cnico,}\\
\it{\small Universidade de Lisboa,
Avenida Rovisco Pais 1, 1049 Lisboa, Portugal}\\
$^c$\,\it{\small Theoretische Physik 1, Naturwissenschaftlich-Technische
    Fakult\"at, Universit\"at Siegen,}\\
\it{\small Walter-Flex-Stra{\ss}e 3, 57068 Siegen, Germany}\\
$^d$\,\it{\small AHEP  Group,  Institut  de  F\'{\i}sica  Corpuscular -- C.S.I.C./Universitat  de  Val\`{e}ncia,}\\
\it{\small Parc  Cient\'{\i}fic  de  Paterna, C/  Catedr\'{a}tico  Jos\'{e}  Beltr\'{a}n,  2  E-46980  Paterna  (Valencia), Spain}\\
}

\maketitle

\begin{abstract}
\noindent  
Motivated by discrete flavour symmetry models, 
we analyse Spontaneous CP Violation (SCPV) for potentials involving 
three or six Higgs fields (both electroweak doublets and singlets)
which fall into irreducible triplet representations of discrete symmetries belonging to the $\Delta(3n^2)$ and $\Delta(6n^2)$ series, including $A_4$, $S_4$, $\Delta(27)$ and $\Delta(54)$.
For each case, we give the potential and find various global mimima for different regions of the parameter space of the potential.
Using CP-odd basis Invariants that indicate the presence of Spontaneous CP Violation we separate the VEVs into those that do or do not violate CP. In cases where CP is preserved we reveal a CP symmetry of the potential that is preserved by those VEVs, otherwise we display a non-zero CP-odd Invariant. 
Finally we identify interesting cases where there is Spontaneous Geometrical CP Violation in which the VEVs have calculable phases.
\end{abstract}

\vfill

\newpage 
\setcounter{page}{2}
\tableofcontents

\section{Introduction}

The discovery of the 
Higgs boson by the Large Hadron Collider (LHC) \cite{Aad:2012tfa,Chatrchyan:2012ufa},
indicates that at least one Higgs doublet must be responsible for electroweak symmetry breaking (EWSB).
However, there is no special reason why there should be only one Higgs doublet in Nature,
and it is entirely possible that there could be additional Higgs doublets, accompanied by further Higgs bosons
which could be discovered in the next run of the LHC. 

The simplest example is two-Higgs-doublet models 
(2HDMs) \cite{Gunion:1989we,Chang:2013ona}. 
However, 2HDMs generally face severe phenomenological problems with flavour changing neutral currents (FCNCs) and 
possible charge breaking vacua, and it is common to consider restricted classes of models controlled
by various symmetries.
In 2HDMs, the full list of possible 
symmetries of the potential is now known \cite{2HDM}.
In 2HDMs these symmetries can be conserved or spontaneously violated after the EWSB, depending on the coefficients of the potential.
Generalising these results to NHDMs is technically difficult, although there has been
some recent progress in this direction \cite{Ivanov:2011ae,Ivanov:2010ww,Ivanov:2010wz,Mohapatra:2011gp}.

The case of three-Higgs-doublet models (3HDMs) is particularly promising for several reasons. 
To begin with, it is the next simplest example beyond 2HDMs which has been exhaustively studied in the
literature. Furthermore, 3HDMs are more tractable than NHDMs, and all possible finite symmetries (but not all continuous ones) have been identified \cite{Ivanov:2012fp}.
Finally, and perhaps most intriguingly, 3HDMs may shed light on the flavour problem, namely the problem
of the origin and nature of the three families of quarks and leptons, including neutrinos, and their 
pattern of masses, mixings and CP violation. 
Typical examples of such models that use discrete symmetries to constrain the structure of mass matrices need several multiplets of scalar fields that also transform under the same symmetry (for reviews, cf.\ \cite{Altarelli:2010gt,Ishimori:2010au,Grimus:2011fk,King:2013eh,King:2014nza,King:2015aea}). Such models provide a motivation to study multiple SM Higgs singlets (sometimes
called ``flavons'' in this context) as well as electroweak doublets. In the context of flavour models it is natural to consider Higgs doublets or singlets which play the role of ``flavons'' and 
form irreducible triplets under some spontaneously broken discrete family symmetry.
Motivated by the above considerations, we shall study CP violating potentials with both 
three and six Higgs doublets and singlets.

CP symmetry, which for a single field is just the combination of particle-antiparticle exchange and space inversion, is presently known to be violated only by the weak interactions involving quarks in the Standard Model (SM)
\cite{Kobayashi:1973fv}. The origin of the observed SM quark CP violation (CPV) is a natural consequence of three generations of quarks whose mixing 
is described by a complex CKM matrix. Although the CKM matrix can be parameterised in different ways, 
it was realised that the amount of CPV in physical processes always depends on a particular weak basis invariant which can be expressed in terms of the quark mass matrices \cite{Jarlskog:1985ht}.
Although CP is automatically conserved by the Higgs potential of the SM, with more than one Higgs doublet
it is possible that the Higgs potential violates CP, providing a new source of CPV \cite{Lee:1973iz}.
We shall be interested in cases of three and six Higgs doublets and singlets, whose potentials are controlled by 
various non-Abelian discrete symmetries which admit irreducible triplet representations. In particular we are interested in
the cases of such potentials which conserve CP explicitly, but where the vacua of such potentials may spontaneously break CP. We shall analyse this problem using basis independent CP-odd invariants for the following reason.

As already mentioned in the context of the CKM matrix, the study of CP is a subtle topic because of the basis dependent nature of the phases which control CP violation. Similar considerations also apply to the phases which
appear in the parameters of the potentials of multiple scalars.
An important tool to assist in determining whether CP is violated or not are basis independent CP-odd invariants (CPIs), whose usefulness has been shown in the SM in addressing CP violation arising from the CKM matrix,
sourced from the Yukawa couplings. The first example of the use of such invariants was the Jarlskog invariant~\cite{Jarlskog:1985ht},
which was reformulated in~\cite{Bernabeu:1986fc} in a form which is generally valid for an arbitrary number of generations. Generalising the invariant approach~\cite{Bernabeu:1986fc} and applying it to fermion sectors of theories with Majorana neutrinos~\cite{Branco:1986gr} or with discrete symmetries~\cite{Branco:2015hea, Branco:2015gna} leads to other relevant CPIs. 
In extensions of the Higgs sector of the SM, the CP violation arising from the 
parameters of the scalar potential can be studied in a similar basis invariant way as
for the quark sector. For example, in the general two Higgs Doublet Model (2HDM)~\cite{Lee:1973iz} (see \cite{Keus:2015hva} for a recent analysis)
a CPI was identified in~\cite{Mendez:1991gp}. More generally, 
applying the invariant approach to scalar potentials has revealed relevant
CPIs~\cite{Lavoura:1994fv, Botella:1994cs, Branco:2005em}, including for the
2HDM~\cite{Davidson:2005cw, Gunion:2005ja}. 
The basic idea is that if CP is conserved then all CPIs vanish (and vice versa).
If any single CPI is non-zero then CP is violated. Finally, CP violating observables all have to be functions of CPIs.

In a recent paper \cite{Varzielas:2016zjc} we considered 
    yet more general Higgs potentials and adopted the powerful method of
    so-called contraction matrices in order to identify and construct new
    non-trivial CPIs,  which we subsequently applied to potentials involving
three or six Higgs fields (which can be either electroweak doublets or
singlets) which form irreducible triplets under a discrete symmetry
\cite{Varzielas:2016zjc}.
Having translated the well-known technique for constructing CPIs 
to diagrams and contraction matrices, we applied this  formalism
to some physically interesting cases which involve three or six Higgs
fields which fall into irreducible triplet representations of discrete symmetries belonging to the $\Delta(3n^2)$ and $\Delta(6n^2)$ series, including $A_4$, $S_4$, $\Delta(27)$ and $\Delta(54)$.
We were mainly interested in the question of explicit CP violation for such Higgs potentials,
although a simple example of spontaneous CP violation was also discussed.
Here we shall be principally concerned with whether those potentials which respect CP can lead to spontaneous CP violation.
In other words we extend our formalism by including also Vacuum Expectation Values (VEVs), obtaining Spontaneous CPIs (SCPIs) that are non-vanishing if CP is spontaneously violated (as considered earlier in~\cite{Lavoura:1994fv, Botella:1994cs}).

The purpose of this paper, then, is to discuss Spontaneous CP Violation (SCPV) for potentials involving 
three or six Higgs
fields which fall into irreducible triplet representations of discrete symmetries belonging to the $\Delta(3n^2)$ and $\Delta(6n^2)$ series, including $A_4$, $S_4$, $\Delta(27)$ and $\Delta(54)$.
These symmetry groups of the potentials considered in this paper are motivated by the fact that all of them are good candidates for discrete flavour symmetries. 
It should be noted that the actual symmetry of the potential can be different from the symmetry group imposed and where this distinction is important, it will be discussed. 
For each case, we write down the potential and find various  
global mimima for different regions of the parameter space of the potential, as recently summarised in \cite{deMedeirosVarzielas:2017glw}.
In principle one could test which CP symmetries are preserved by VEVs, but it can be non-trivial to know all CP symmetries of the potential, we therefore prefer to use invariants.
In each case we shall consider 
CP-odd basis Invariants (CPIs) that indicate SCPV - which we refer to as Spontaneous CP-odd Invariants (SCPIs). This builds on and was enabled by \cite{Varzielas:2016zjc}, where diagrammatic methods for constructing such invariants were further developed to the point of making them useful for the analysis of complicated example models such as the authors have in mind for this paper.
In cases where CP is preserved we give a CP symmetry of the potential that is preserved by those VEVs, otherwise we show a non-zero CP-odd Invariant. 
In models where CP is violated spontaneously, thanks to the enhanced symmetry at high energies, the number of parameters of the model can be greatly reduced, and thanks to the controlled breaking of CP, the strength of CPV will have to be a function of these parameters too, which relates observable low-energy phenomena to possibly extremely high-energy parameters. 

We emphasise that the work here extends the scope of the existing 
main models of SCPV considered in the literature based on models with 2 Higgs doublets (2HDM) or 3 Higgs doublets (3HDM). We remark that the most general 2HDM has a sufficient number of complex parameters to allow for explicit CPV. If a CP symmetry is imposed on the Lagrangian in the unbroken phase, the explicit CPV disappears and spontaneous CPV becomes possible. However, in both of those cases, when coupled to fermions, the model gives rise to flavour-changing neutral currents. When the latter are forbidden using any kind of symmetry, also both of explicit CPV (i.e.\ CPV with zero VEV) and SCPV disappear. 
3HDMs improve a little on the situation of the 2HDM, as both of  explicit CPV and FCNCs can be eliminated separately by CP and flavour-type symmetries, such that SCPV is possible without FCNCs. 
In particular a general analysis of CP invariants in 3HDMs suitable for SCPV does not exist, and 
one of the motivations of this work is to extend the discussion of CPIs suitable for these cases.
We are also interested in models with 6 Higgs doublets (6HDM), for which the question of SCPV has not been analysed at all, and we extend our analysis to such cases also.
Specifically, we expand our analysis by considering also the 6-Higgs potentials invariant under the discrete symmetries $A_4$, $S_4$, $\Delta(27)$, $\Delta(54)$, $\Delta(3n^2)$ and $\Delta(6n^2)$ with $n>3$.
In very special symmetric cases a phenomenon called geometric CP violation (GCPV) arises which means that values of complex CP-violating phases of VEVs do no longer depend on the parameters of the potentials in the region of the parameter space where they are minima. The only 
case 
that was known for a long time was that of Higgs triplets of $\Delta(27)$ \cite{Branco:1983tn}. Recently, several other cases had been discussed by \cite{Varzielas:2012pd, Ivanov:2013nla}. We shall discuss new examples of such GCPV.

It is important to further clarify which relevant results are already available in the literature, and which we need to
obtain for the first time.
A complete list of possible global minima for the $A_4$, $S_4$, $\Delta(27)$ and $\Delta(54)$ Higgs potentials with 3 Higgs fields has been obtained by \cite{Ivanov:2014doa}, which includes furthermore cases without irreducible triplets such as $S_3$ (the $S_3$ is itself studied in great detail in \cite{Emmanuel-Costa:2016vej}).
For the cases where the respective potential is CP conserving in general (e.g.\ $A_4$) \cite{Ivanov:2014doa} also verified none of these VEVs spontaneously violate CP. While we analyse cases with 3 Higgs arranged as irreducible triplets, we go beyond confirming the existing results for the 3-Higgs potentials with SCPIs, employing our methodology to analyse the VEVs and CP properties of other relevant cases.
We continue by checking 3-Higgs potentials where a specific CP symmetry is imposed in addition to a discrete symmetry for which the potential is in general CP violating (e.g.\ the $\Delta(27)$, $\Delta(54)$) - this includes the case where Spontaneous Geometrical CP Violation (SGCPV) was first identified \cite{Branco:1983tn}.

The layout of the paper is as follows: In Section \ref{sec:CPV} we set our notation and describe how to identify CP violation and spontaneous CP violation within the basis invariant formalism. In Section \ref{sec:SCPI} we present the spontaneous CP invariants (SCPIs) that we will use throughout the paper. In Section \ref{sec:VEVs} we list the discrete symmetry groups and the potentials invariant under them that we consider (with one or two triplets), as well as a list of candidate Vacuum Expectation Values (VEVs) that are global minima of some of those potentials. In Sections \ref{sec:onetrip}, and \ref{sec:twotrip} we apply the SCPIs to the potentials invariant under the respective symmetry groups, checking if CP is conserved when all SCPIs we calculate vanish, and otherwise checking which of the VEVs we have found spontaneously violate CP. For any VEVs that do violate CP, we further consider if their phases are calculable, i.e.\ if there is Spontaneous Geometrical CP Violation. Section \ref{sec:sum} is a summary of our results and Section \ref{sec:con} concludes the paper.

\newpage
\section{CP violation \label{sec:CPV}}
In this section we review the current understanding of CP symmetries in models of several scalar fields.
\subsection{Generalities}
Usually, CP transformations are thought to act on scalars via complex conjugation of the field itself and parity transformation of the field coordinates, combined with arbitrary unitary basis transformations between fields with equal quantum numbers to form what is then called generalized CP transformations. A multiplet of scalar fields $\varphi=(\varphi_1,\ldots,\varphi_n)$ would then be to transform under such a generalized CP transformation as
\begin{equation}
 \varphi \mapsto X \varphi^\ast (x^P)
\end{equation}
with a unitary matrix $X$ and $x^P=(t, -{\bf x})$. Recently, \cite{Aranda:2016qmp} re-opened the discussion about how for complex scalars that have no charge associated to an Abelian symmetry, CP transformations without complex conjugation are equivalent to such with complex conjugation, by which only the (generalised) parity part of the transformation plays a role.

In the scalar potentials considered by us, fields are assumed, at least before symmetry breaking, to carry some conserved $U(1)$ charge, as happens automatically for Higgs doublets, and imposed on EW singlets where they are considered, in the latter case primarily to render the potential even. For this reason, we think that at least symmetries of CP-type before the symmetry breaking will involve complex conjugation for the fields considered here.

When defining a model, in principle arbitrary CP-type symmetries can just be imposed onto a model at high energies in addition to all its other symmetries to render it explicitly CP-conserving and in addition maybe constrain it in other ways. If the potential has both pure flavour-type and CP-type symmetries, the notion of the \textit{consistency} of these symmetries exists.

By this one means that only certain CP-type symmetries can be imposed onto a model without enlarging the flavour-type symmetries of the potential. In essence, apart from the different physical interpretation, there is no difference between flavour-type and CP transformations, as both just relate different degrees of freedom. Consider a set of scalar fields $\varphi$ and its complex conjugate, and combine them as
\begin{equation}
 \phi=\begin{pmatrix}\varphi\\\varphi^\ast\end{pmatrix}.
\end{equation}
When written in this way, pure flavour-type and CP-type transformations act in the following ways:
\begin{equation}
 \phi\xrightarrow{flavour} \begin{pmatrix}\rho&0\\0&\rho^\ast\end{pmatrix}\phi~\text{, and}~ \phi\xrightarrow{CP}\begin{pmatrix}0&X\\X^\ast&0\end{pmatrix}\phi.
\end{equation}
Next, consider basis transformations $U$ that at this stage (before symmetry breaking) only connect fields with identical (gauge) charges and thus act in the same way as a flavour transformation on the fields. There are some basis transformations for which $U\rho U^\dagger$ is just another group element $\rho'$. If the potential already has some CP symmetries, then such basis transformations generate additional CP transformations without changing the flavour type symmetry at all
(and can be eliminated in the same way). The CP symmetries generated in this way are of the form
\begin{equation}
 \begin{pmatrix}0&UXU^T\\U^\ast X^\ast U^\dagger&0\end{pmatrix}.
\end{equation}
Such basis transformations had been considered in \cite{Fallbacher:2015rea} and have the effect of leaving the potential form-invariant and relate different points in the parameter space of the potential.

However, an arbitrary CP transformation with matrix $X$, which splits into symmetric and antisymmetric parts, $X=X_s+X_a$, transforms under a flavour basis transformation as follows,
\begin{equation}
 UXU^T=UX_sU^T+UX_aU^T=\tilde X_s+UX_aU^T=: \tilde X,
 \label{CP_basis_trafo}
\end{equation}
where in the second step $U$ can now be chosen such that it diagonalises $X_s$ while $UX_aU^T$ is still antisymmetric. Thus, always a flavour basis exists where at least for one of the pure CP-type symmetries that are not related by flavour transformations or transformations that leave the potential form-invariant the corresponding $X$ can be made the sum of a diagonal and an antisymmetric matrix. In addition, if $X$ was symmetric from the start, it can be made the identity matrix. This basis transformation does not change the size of the overall symmetry of the potential which is generated by all of the above $2\times2$ block matrices corresponding to flavour and CP transformations. Furthermore, this basis transformation generally does not leave the flavour symmetry or the potential form-invariant.
The order-4 CP symmetry discussed in \cite{Ivanov:2015mwl} and \cite{Aranda:2016qmp} is of this form of a sum of a diagonal and an antisymmetric matrix.

Arbitrary CP transformations can enlarge the pure flavour-type symmetry. The simplest way in which imposed flavour and CP symmetries combine to a pure flavour-transformation is
\begin{equation}
 \begin{pmatrix}0&X\\X^\ast&0\end{pmatrix}\begin{pmatrix}\rho&0\\0&\rho^\ast\end{pmatrix}\begin{pmatrix}0&X\\X^\ast&0\end{pmatrix}=\begin{pmatrix}X \rho^\ast X^\ast&0\\0&X^\ast \rho X\end{pmatrix}
\end{equation}
If now $X\rho^\ast X^\ast$ is not another element of the imposed flavour group, the actual flavour-type symmetry has been enlarged. 

In addition, it is clear that even powers of CP transformations are flavour transformations and corresponding conditions arise for $(X X^\ast)^n$.

Next, if the imposed $X$ was symmetric and one can use a basis transformation to go to a basis where $\tilde X=1$, combining CP and flavour transformations as above gives
\begin{align}
 \begin{pmatrix}0&X\\X^\ast&0\end{pmatrix}\begin{pmatrix}\rho&0\\0&\rho^\ast\end{pmatrix}\begin{pmatrix}0&X\\X^\ast&0\end{pmatrix}&\rightarrow\begin{pmatrix}0&1\\1&0\end{pmatrix}\begin{pmatrix}U&0\\0&U^\ast\end{pmatrix}\begin{pmatrix}\rho&0\\0&\rho^\ast\end{pmatrix}\begin{pmatrix}U^\dagger&0\\0&U^T\end{pmatrix}\begin{pmatrix}0&1\\1&0\end{pmatrix}\nonumber\\
&=\begin{pmatrix}U^\ast \rho^\ast U^T&0\\0&U\rho U^\dagger\end{pmatrix} 
\end{align}
while the original flavour transformations become $U \rho U^\dagger$ and the flavour symmetry in the basis where $\tilde X=1$ is generated by all of $U \rho U^\dagger$ and of $U^\ast \rho^\ast U^T$. On the other hand, when combined with CP with $X=1$, the complete original flavour symmetry was generated by all of $\rho$ and of $\rho^\ast$.

For CP transformations where no basis exists where $X$ can be made the identity, the same conditions arise, only with an additional matrix $\tilde X$ appearing, and the full flavour symmetry in the basis where the symmetric part of $X$ has been made diagonal is generated by all of  $U \rho U^\dagger$ and of $\tilde X U^\ast \rho^\ast U^T \tilde X^\ast$.

Conversely, when one assumes that the flavour symmetry is complete and the CP part of the full symmetry is generated by a single matrix $X$, and as observables of CPV are invariants under internal basis transformations, this means that \textit{for fixed flavour-type symmetry combined with a single CP generator, the physically different CP generators are of the form of the RHS of Eq.~(\ref{CP_basis_trafo})} (up to changes of the basis of the group matrices that does not extend the flavour-type symmetry).

For models with more than one CP generator, the situation can be more complicated, as in the basis where one of them is of the form of Eq.~(\ref{CP_basis_trafo}), the other can still be arbitrary. Unfortunately, this situation has to be outside the scope of this paper.

Later, when example potentials are studied, for chosen flavour symmetries, we still impose various CP symmetries and boldly ignore the question of consistency. These CP symmetries will then enlarge the actual flavour-type symmetry of the potential, and building a model this way is still fine, if one is aware that the full symmetry is always generated by the $2\times2$ flavour-type and CP-type block matrices. The first potential where this is relevant has at least initially a $\Delta(54)$ flavour symmetry and further discussion can be found in the respective section.

\subsection{Spontaneous CP Violation}
\label{spont_CPV_subsec}
One of the questions we want to investigate in this paper is, when does geometrical CPV arise spontaneously (SGCPV) --- what are the conditions on the potential or on the symmetries of the potential? Finding more cases with SGCPV is a step in this direction. In the literature, e.g.\ \cite{Varzielas:2012pd, Ivanov:2013nla}, geometrical CP violation has been defined as the situation when the relative phases of a VEV become ``calculable'' which is to mean that in the region of parameter space of the potential where this VEV is a (global) minimum, these relative phases do not depend on the model parameters. This criterion has two components, one, that the VEV is geometric, and second, that CP is violated spontaneously by it. Geometric CPV is theoretically appealing because the strength of CP violation is no longer a function simply of arbitrary and at least in the near future unmeasurable parameters and furthermore also as phases arising in geometric CPV are stable against renormalization, as they are protected by a residual symmetry \cite{Branco:1983tn}.

One good criterion for spontaneous violation of CP that we rely on throughout the paper is when at least one CP-odd invariant that can indicate spontaneous CPV (see next subsection) is non-zero. When CP is violated, then at least one CP-odd invariant is non-zero, but it is not possible at the moment to obtain a complete list of invariants and it is impractical to test all of them in any case. The situation would be simpler, if one had a basis of CP-odd invariants, which is a (small) finite set of CP-odd invariants with the property that if all of them are zero, CP is conserved. Indeed, such a basis of CP-odd invariants is known for the 2HDM, \cite{Gunion:2005ja}, but for the more symmetric potentials we will consider, all invariants from \cite{Gunion:2005ja} vanish trivially without indicating CP-conservation, because whether a set of invariants forms a basis is model-dependent, which is why in \cite{Varzielas:2016zjc} it was found necessary to find additional CP-odd invariants. Finally, if CP is conserved, then all CP-odd invariants (which includes all possible basis sets) vanish of course. Again, the strength of CP-odd invariants that indicate spontaneous CP violation is that it is not necessary to systematically know all CP symmetries of the unbroken potential, a task which can be non-trivial. 

Without SCPIs, one could also check directly whether a VEV preserves a specific CP symmetry that leaves the potential invariant. This is arguably more direct, but it has the significant drawback that it is not sufficient to find a CP symmetry not preserved by the VEV for there to be CPV, but rather all CP symmetries that left the potential invariant must be broken by the VEV. Finding all inequivalent CP symmetries that leave a potential invariant can be non-trivial, as the above discussion elaborates, and checking each is broken by the particular VEV is also cumbersome. Conversely, if a single CP symmetry that leaves the (unbroken) potential invariant is preserved by the VEV, then CP is conserved (this is a sufficient condition). This makes the direct check very convenient to confirm CP is preserved. We now review this direct condition in a notation that follows \cite{Varzielas:2016zjc}. Consider an even scalar potential written in the following standard form,
\begin{equation}
 V(\varphi)=Y^a_b \varphi_a \varphi^{\ast b}+Z^{ab}_{cd}\varphi_{a}\varphi_{b}\varphi^{\ast c}\varphi^{\ast d},
 \label{V_standard_form}
\end{equation}
where $\varphi$ contains as components the components of all fields. $Y$ and $Z$ are tensors that contain all allowed couplings and are subject to possible symmetries acting on $\varphi$. In the following only the action of CP transformations is repeated from \cite{Varzielas:2016zjc}.\footnote{
The rules for indices can be summarized as follows: On $\varphi$, $ (\varphi_a)^\ast=\varphi^{\ast a},~(\varphi^{\ast a})^\ast=\varphi_a$ and on matrices, e.g.\ a basis transformation $U$, $({U_a}^b)^\ast={U^{\ast a}}_b$. Consequently, $({U_a}^b\varphi_b)\ast={U^{\ast a}}_b \varphi^{\ast b}={(U^\dagger)_b}^a \varphi^{\ast b}$.}
Assume that this potential is CP-conserving, so in particular invariant under a set of CP transformations of the fields, such that when one transforms the fields as in
\begin{equation}
 \varphi\mapsto X \varphi^\ast,
 \label{CP_trafo_definition}
\end{equation}
with a unitary matrix $X$, the potential is unchanged,
\begin{equation}
 V(\varphi)=V(X\varphi^\ast).
\end{equation}
With indices explicitly shown, we write the transformation of Eq.\ (\ref{CP_trafo_definition}) as
\footnote{This is a slight improvement on the notation in \cite{Varzielas:2016zjc}, where CP transformations were written with one upper and one lower index.}
\begin{equation}
 \varphi_a\mapsto X_{a a'} \varphi^{\ast a'}:=\sum_{a'}X_{a a'} (\varphi_{a'})^\ast
 \label{CP_trafo_indices_definition}
\end{equation}
and for the complex conjugated field as
\begin{equation}
 \varphi^{\ast a}\mapsto X^{\ast a a'}\varphi_{a'}.
 \label{CP_trafo_conjugated}
\end{equation}
Note that in this notation, $X^{\ast a a'}=X^{\dagger a'a}$.
Denote the whole set of such $X$ under which a potential is invariant in the unbroken phase by $\mathcal{X}:=\{X\}$.
In Appendix \ref{technical_details_appendix} the traditional argument is repeated in the notation of Eq.~(\ref{V_standard_form}) that a VEV $v=\langle\varphi\rangle$ conserves CP if at least for one of the matrices $X\in \mathcal{X}$ holds that
\begin{equation}
 v_{a}=X_{aa'}v^{\ast a'}.
 \label{VEV_CP_conservation_new}
\end{equation}
Inverting the argument this means that a VEV violates CP if for none of the original CP trafos in the unbroken phase, Eqs.~(\ref{CP_trafo_indices_definition}), (\ref{CP_trafo_conjugated}), the previous condition can be fulfilled. 

Recently, in \cite{Ratz:2016scn}, it was argued that CP could be violated while a CP transformation in the broken phase can be found. The reason why this is not in contradiction to the previous argument is that the CP symmetry in the broken phase found in \cite{Ratz:2016scn} was not a symmetry of the unbroken phase. Furthermore, the type of CP transformations without complex conjugation discussed in \cite{Aranda:2016qmp}, while they might be a symmetry of the neutral components of scalar fields after symmetry breaking, are in our case also not symmetries in the unbroken phase. 

\subsection{Spontaneous CP Invariants \label{sec:SCPI}}

From the standard form of the potential, Eq.~(\ref{V_standard_form}), invariants can be built in a model-independent way that are CP-odd by construction. Model-independent here is to mean that these invariants are constructed from the $Y$ and $Z$ tensor in Eq.~(\ref{V_standard_form}) and work for every potential that can be put into this standard form.

Under basis transformations, the field multiplet and its conjugate transform as $\varphi \mapsto U \varphi$ with which the $Y$ and $Z$ tensor of the even potential in the standard form of Eq.~(\ref{V_standard_form}) transform under basis transformations of the fields and its conjugate as
\begin{align}
Y_a^b &\mapsto  U_a^{a'} \, Y_{a'}^{b'} \, {U^\dagger}_{b'}^b \ ,\\
Z_{ac}^{bd} &\mapsto  U_a^{a'}\, U_c^{c'} \, Z_{a'c'}^{b'd'} \, 
{U^\dagger}_{b'}^b \,{U^\dagger}_{d'}^d \ .
\end{align}
Additionally, VEVs transform as vectors under basis transformations,
\begin{equation}
 v_a \mapsto U_a^{a'}v_{a'}\text{ and } v^{a^\ast}\mapsto U^a_{a'}v^{a'\ast}
\end{equation}
and one can see now that every combination of $Y$, $Z$, $v$ and $v^\ast$ where all indices are correctly contracted forms a basis invariant. Complex conjugation acts on invariants by interchanging upper and lower indices, such that the complex conjugate of a basis invariant can be obtained by interchanging all upper and lower indices of all components. In \cite{Varzielas:2016zjc}, invariants had been related to diagrams where arrows indicate index contractions and other parts of the diagram tensors and VEVs. In that formalism, an invariant is CP-odd if the diagram does not stay identical when inverting the direction of all arrows, which is the diagrammatic equivalent to complex conjugation. The diagrammatic formalism furthermore allowed for a systematic search for CP-odd invariants. In \cite{Varzielas:2016zjc}, then all inequivalent CP-odd invariants up to some order had been constructed, especially also such that indicate spontaneous CPV.

\begin{figure}[h]
\begin{center}
\includegraphics[scale=0.22]{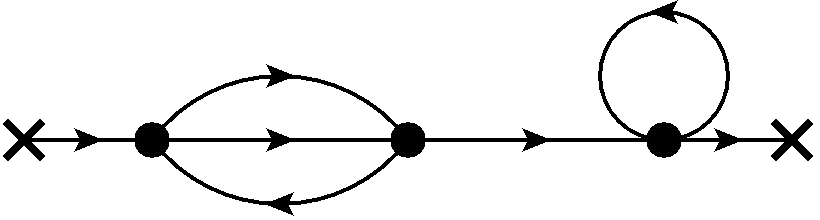}
\end{center}
\caption{The diagram corresponding to the invariant Eq.~(\ref{ZZZvv}). (In \cite{Varzielas:2016zjc}, the precise relation between invariants and diagrams is explained.}
\label{ZZZvv_diagram}
\end{figure}

All invariants for SCPV (SCPIs) that were listed in appendix B.5 of \cite{Varzielas:2016zjc} have been evaluated for all example potentials that will be considered in later sections. 
As a first example, we consider here briefly the most general 2-Higgs Doublet Model (2HDM) potential.
In our notation, brackets indicate the $SU(2)_L$ contractions e.g.\ $(H_1^\dagger H_1)^2 = (h_{1,1}^\dagger h_{1,1} + h_{1,2}^\dagger h_{1,2})^2$, therefore we write the general two-Higgs-doublet (2HDM) potential as
\begin{eqnarray}
V (H_{1},H_{2})&=&m_{1}^2 \ H _{1}^{\dagger }H _{1}+ m_{12}^2\ e^{i\theta_0 }\ H
_{1}^{\dagger }H _{2}+ m_{12}^2 \ e^{-i\theta_0 }\ \ H _{2}^{\dagger }H
_{1}+m_{2}^2\ H _{2}^{\dagger }H _{2}+ \nonumber \\[2mm]
&&+a_{1}\ \left( H _{1}^{\dagger }H _{1}\right) ^{2}+a_{2}\ \left( H
_{2}^{\dagger }H _{2}\right) ^{2}\nonumber \\[2mm]
&&+b\ \left( H _{1}^{\dagger }H
_{1}\right) \left( H _{2}^{\dagger }H _{2}\right) +b^{\prime }\ \left(
H _{1}^{\dagger }H _{2}\right) \left( H _{2}^{\dagger }H
_{1}\right) + \nonumber \\[2mm]
&&+c_{1}\ e^{i\theta _{1}}\ \left( H _{1}^{\dagger }H _{1}\right) \left(
H _{2}^{\dagger }H _{1}\right) +c_{1}\ e^{-i\theta _{1}}\ \left( H
_{1}^{\dagger }H _{1}\right) \left( H _{1}^{\dagger }H _{2}\right) +
\nonumber \\[2mm] 
&&+c_{2}\ e^{i\theta _{2}}\ \left( H _{2}^{\dagger }H _{2}\right) \left(
H _{2}^{\dagger }H _{1}\right) +c_{2}\ e^{-i\theta _{2}}\ \left( H
_{2}^{\dagger }H _{2}\right) \left( H _{1}^{\dagger }H _{2}\right) +
\nonumber \\[2mm]
&&+d\ e^{i\theta_3 }\ \left( H _{1}^{\dagger }H _{2}\right) ^{2}+d\
e^{-i\theta_3 }\ \left( H _{2}^{\dagger }H _{1}\right) ^{2}.
\label{2SU2}
\end{eqnarray}
In it appear $H_1 = (h_{1,1},h_{1,2})$ and $H_2 = (h_{2,1},h_{2,2})$, and the arbitrary coefficients $a_1$, $a_2$, $b$, $b'$,$c_1$, $c_2$, $d$.
For the 2HDM, all SCPIs in \cite{Varzielas:2016zjc} are non-zero although we don't show any as even the smallest expression is rather large. In any case they are not completely meaningful at this stage, as the general 2HDM is explicitly CP violating. When imposing a CP symmetry, the expression becomes meaningful - for example when imposing trivial CP, the expression for the CP-odd invariant
\begin{equation}
 \mathcal{J}_{2HDM}\equiv Z^{a_1a_2}_{a_3a_7}Z^{a_3a_4}_{a_1a_2}Z^{a_5a_6}_{a_4a_6}v_5 v^{\ast a_7}-c.c.,
 \label{ZZZvv}
\end{equation}
which of the invariants tested produces the smallest non-zero expressions for the 2HDM, becomes a function of the coefficients in the potential multiplying the VEVs in the following combination:
\begin{equation}
\mathcal{J}_{2HDM}=F(a_1, a_2, b, b',c_1,c_2, d)
[\langle h_{2,1} \rangle \langle h_{1,1} \rangle^*+ \langle h_{2,2} \rangle \langle h_{1,2} \rangle^* 
- h.c. ]
\end{equation}
i.e.\ it correctly identifies that SCPV depends in this case on the relative phase between the VEVs of $H_1$ and $H_2$ (note that charge preserving VEVs correspond to $\langle h_{1,1} \rangle = \langle h_{2,1} \rangle = 0$). The diagram corresponding to this invariant is shown in Figure \ref{ZZZvv_diagram}. The same quadratic VEV dependence factors out also when calculating the $J_{1}^{(3,2)}$ SCPI that has 2 pairs of VEVs, as the full expression is a more complicated quadratic function of the VEVs multiplying $[\langle h_{2,1} \rangle \langle h_{1,1} \rangle^*+ \langle h_{2,2} \rangle \langle h_{1,2} \rangle^*
- h.c. ]$.
For the more symmetric potentials considered in the remainder, the above simple invariant vanishes trivially, i.e.\ without indicating CP-conservation.
We will not consider the 2HDM further and will instead use as our examples 3HDM and 6HDM potentials that are invariant under discrete symmetries.

\begin{figure}[h]
\begin{center}
\includegraphics[scale=0.22]{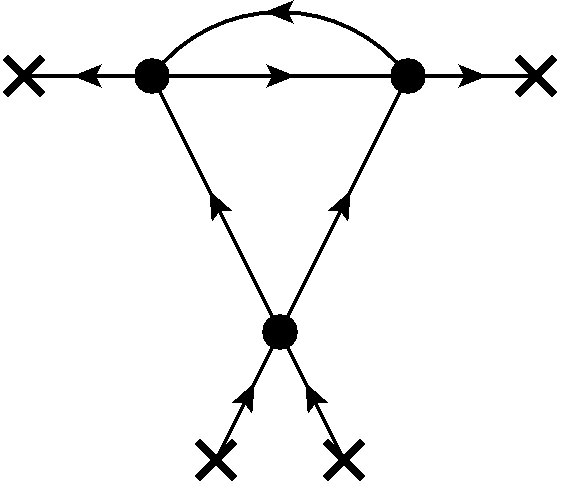}
\end{center}
\caption{The diagram corresponding to the invariant $J^{(3,2)}$.}
\label{J321_diagram}
\end{figure}

\begin{figure}[h]
\begin{center}
\includegraphics[scale=1]{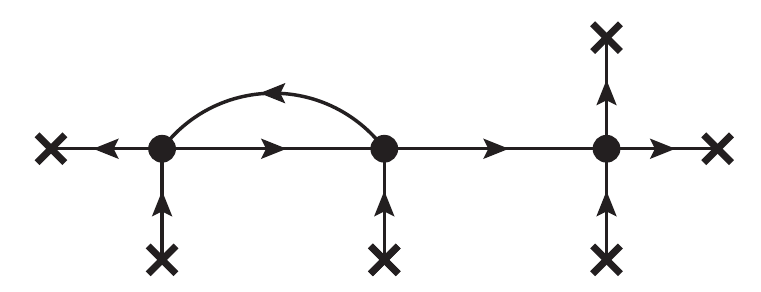}
\end{center}
\caption{The diagram corresponding to the invariant $J^{(3,3)}$.}
\label{ZZZv6_diagram}
\end{figure}

We recall that SCPIs are obtained by subtracting contractions of tensors to make them CP-odd, $\mathcal J = J - J^*$.
For all potentials besides the 2HDM that we consider as examples, of the SCPIs listed in \cite{Varzielas:2016zjc} only $\mathcal J_{1}^{(3,2)}$ is non-zero. For this reason we drop the subscript, referring to it as $\mathcal J^{(3,2)}$ (we keep the superscript to distinguish from another SCPI which we will use). The contraction $J^{(3,2)}$ of $Z$ tensors and VEVs $v$ out of which $\mathcal J^{(3,2)}$ is formed, is in terms of the $Z$ tensor and VEVs given by
\begin{equation}
J^{(3,2)}\equiv Z^{a_1a_2}_{a_4a_5}Z^{a_3a_4}_{a_2a_6}Z^{a_5a_6}_{a_7a_8}v_{a_1}v_{a_3}v^{\ast a_7}v^{\ast a_8}.
\end{equation}
The diagram corresponding to this invariant is shown in Figure \ref{J321_diagram}, where it also be seen that it is CP-odd as the diagram with inverted arrows cannot be made identical to the original diagram, however the dots (= $Z$ tensors) and crosses (= VEVs) are moved around.
Given that only one of the previously constructed SCPIs is non-trivial for the potentials we are interested in, the question arises if there is another SCPI which gives
non-trivial results. Searching for SCPIs with an additional $Z$ tensor, i.e.\ considering invariants of type $J^{(4,2)}_i$, an
explicit (but not systematic) search did not yield a positive
answer. On the other hand, an invariant $J^{(3,3)}$ with an additional pair
of VEVs revealed another interesting SCPI which turned out to be sufficient for our purposes (and therefore we also do not use a subscript to label it):
\be
J^{(3,3)}\equiv Z_{a_1a_2}^{a_7a_8}Z_{a_3a_4}^{a_5a_9}Z_{a_5a_6}^{a_1a_3}
v_{a_7}v_{a_8}v_{a_9}
v^{\ast a_2}v^{\ast a_4}v^{\ast a_6}\ .
\ee
The diagram corresponding to this invariant is shown in Figure \ref{ZZZv6_diagram}.

The results of evaluating both these SCPIs for the various potentials, CP symmetries and VEVs is summarized in Section \ref{sec:sum}, including Table~\ref{ta:1_summary} and Table~\ref{ta:2_summary}.
As shown in the following sections where different potentials are considered in detail, $J_{1}^{(3,3)}$ provides us with additional relevant information that complements what we learn from $\mathcal J^{(3,2)}$. For example, from our examples, the simplest 3 field potential that for arbitrary VEV values (which not necessarily can be obtained from the potential) gives a non-vanishing expression for $\mathcal  J_{1}^{(3,3)}$ is invariant under $S_4$, $V_{S_4} (\varphi)$. 
The SCPI $J^{(3,2)}$ gives a vanishing expression independently of the VEVs considered, whereas the expression for $\mathcal  J^{(3,3)}$ is in general:
\bea
\mathcal  J^{(3,3)} &=& c (c^2-s^2) \times \nonumber\\
&& \Big[
|v_1|^2 v_1^2\big( v_2^{\ast 2} +  v_3^{\ast 2} \big) + 
|v_2|^2 v_2^2\big( v_3^{\ast 2} +  v_1^{\ast 2} \big) + 
|v_3|^2 v_3^2\big( v_1^{\ast 2} +  v_2^{\ast 2} \big) 
~-~ \mathrm{h.c.}\Big]  .~~~~~
\eea
Note that this doesn't mean that it is possible to have SCPV in this potential, as when plugging in any of the possible VEVs \cite{Ivanov:2014doa}, $\mathcal  J^{(3,3)}$ does indeed vanish, as shown in Section \ref{sec:S_A4} and summarised in Section \ref{sec:sum}.

\section{Potentials and Vacuum Expectation Values \label{sec:VEVs}}
\cleqn

In this paper we are mainly interested in three- and six-Higgs-doublet models,
where fields form one or two irreducible triplet
representations of some discrete symmetry group. 
The most complicated potentials that will be considered in this paper then describe six scalar $SU(2)_L$ doublets (or if only $U(1)_{em}$-preserving minima are considered, equivalently, $SU(2)_L$ singlets), and finding (global) minima can be a non-trivial task.

The aim of this section is to arrive at a list (not necessarily exhaustive, but complete enough for further analysis) of possible VEVs of potentials of two scalar triplets of $A_4$, $S_4$, $\Delta(27)$, $\Delta(54)$, and $\Delta(3n^2)$ and $\Delta(6n^2)$ with $n>3$. 

First, the potentials of one triplet will be re-analysed, after which the potentials of two triplets are examined.
As the aim in the next section will be to study geometrical CPV, potentials are made CP-conserving via various additional CP symmetries.

We identify classes of potentials by the symmetry imposed by us. This differs from \cite{Ivanov:2014doa}, where potentials were identified by their full symmetry, both of flavour- and CP-type, which was enabled by a classification of those symmetries in \cite{Ivanov:2012fp}. The full symmetry of the potential can be different from the imposed symmetry, e.g.\ because 
cutting off the potential at renormalisable operators can enlarge the symmetry group. 

The full symmetry of the potential or of a part of it is what will be relevant in the following, and it will hopefully always be made clear which exact symmetry is under discussion.

For $SU(2)_L$ singlets, always an additional $U(1)$ symmetry was imposed to force the potential to be even. In addition, to not break $U(1)_{em}$, in a basis where at least one VEV component of one doublet is zero, also the same component of the other doublets has to be zero. When in a term in a potential of singlet fields the flavour indices allow for various ways of contracting $SU(2)_L$ indices, then if without loss of generality, the bottom component of all doublets is set to zero, such terms in doublet potentials will coincide again.

For the aforementioned reasons the possible vacuum expectation values that don't break $U(1)_{em}$ of a doublet potential with some symmetry will be identical to those of the corresponding singlet potential made even by an additional $U(1)$.
Some of these potentials had been considered in \cite{Ivanov:2014doa}, where also the full list of global VEVs was obtained via a geometric method.

When searching for the minima of a potential, one is not just interested in the minimum at some position in the parameter space of the potential, but one would ideally like to know what the minima of the potential are in every corner of its parameter space.

Minima that are related by symmetries of the potential have the same energy by definition and such sets of related VEVs are called orbits, of which a potential can have more than one. Notably, when the symmetry of the potential is increased (of flavour- or CP-type), this has the effect of merging orbits. 
Beyond this is should be noted that the full symmetry of the VEV set is larger than that of the potential. The set of VEVs is also invariant under transformations under which the potential was just form-invariant, cf.\ \cite{Fallbacher:2015rea}, as such transformations connect orbits.
In other words, while the symmetries of the potential are valid at every point of the whole space of the parameters of the potential, the set of VEVs additionally contains information about all potentials that differ by the values of their parameters but not by their symmetries.

In the notation of section \ref{spont_CPV_subsec}, classical minima of the potential are solutions of
\begin{equation}
 0=\frac{\partial V}{\partial v_i}=Y^i_b v^{\ast b}+2 Z^{ib}_{cd}v_v v^{\ast c}v^{\ast d}
 \label{first_Ds}
\end{equation}
and the corresponding conjugated equation, as well as that the matrix formed out of second derivatives is positive definite. 
The mixed derivative by VEV and conjugated VEV would e.g.\ be
\begin{equation}
\frac{\partial^2V}{\partial v_i \partial v^{\ast j}}=Y^i_j+4Z^{ib}_{jd}v_b v^{\ast d}.
\label{second_Ds}
\end{equation}

An alignment is basically, in accord with the literature, a VEV where one does not care about its absolute length but only about its direction. The linear symmetries in consideration can in any case not constrain the length of the VEV, as the overall sizes of quadratic and quartic terms are entirely unrelated. At the end of the day one is interested in interesting alignments, especially such that violate CP spontaneously or even geometrically. In a following section, CP violation will be discussed in greater detail, this section focuses on obtaining VEVs of the potentials considered in this paper. 

New VEVs are generated by analysing which degrees of freedom become physical when going from from a potential with high symmetry to one with lower symmetry. Consider a potential 
\begin{equation}
V=V_G+V_H\text{ with } H\subset G.
\end{equation}
All alignments $v_i$ of potentials generally fall into orbits  $\{H\cdot v_i\}$. If one considers the subpotential $V_G$ separately, it will have minima that again fall into orbits $\{G\cdot v_i\}$. Next, consider $V_H$ as terms that explicitly break the symmetry of the potential from $G$ to $H$. By this, the orbits of $V_G$ are now split into several orbits organized by $H$, sometimes only a finite number of orbits, occasionally there can be a new continuous parameter distinguishing them. In this case the full potential can be minimized in this new parameter, which is often possible analytically. 

First, it is thus  tested for potentials with one triplet of scalars, which orbits can be obtained from the orbits of more symmetric potentials, just by applying the symmetries of the less symmetric potential to the orbits of the more symmetric one. In this way, degrees of freedom that were unphysical in the more symmetric potential can become physical and after minimizing just the part of the potential that depends on these new degrees of freedom, one obtains as global minima exactly those that had also been obtained in \cite{Ivanov:2014doa}. One could thus conjecture that for certain relations between symmetry groups, also a relation between the global minima of potentials and these symmetries exists which might be made more precise.

After that, the same method is applied to obtain some of the (conjectured) global minima of potentials of two triplets of scalars from those of potentials that are just the sum of potentials of one triplet without cross-terms. Again this simply shows which degrees of freedom become physical due to the reduced symmetry. Still, many new interesting global minima are obtained, many of which are geometric and many of which will turn out to be CP violating and especially geometrically CP violating, which constitute the first examples of this phenomenon outside the original $\Delta(27)$ model and \cite{Varzielas:2012pd, Ivanov:2013nla}. The cases found here are in particular the first new examples of geometric CPV with six scalars.

Generally, a VEV is defined as being geometric when in the region of the parameter space of the potential where it is a (global) minimum,  its direction but not its length depend on the potential parameters, or if the potential is expressed in terms of $Y$ and $Z$ tensors, 
\begin{equation}
 v=|v|(Y,Z) \hat v,
\end{equation}
where $|v|$ is the length of the VEV and the only part that depends on the potential parameters, and $\hat v$ is the normalised direction that does not depend on them. This is equivalent to 
\begin{equation}
 \frac{\partial \hat v}{\partial Y^a_b}=0 \text{ and }\frac{\partial \hat v}{\partial Z^{ab}_{cd}}=0 .
\end{equation}
A VEV is primarily a solution of the minimisation conditions, in the notation of Eq.~(\ref{V_standard_form}), given by
\begin{equation}
 \frac{\partial V}{\partial \phi_i}\rightarrow Y^i_b v^{\ast}+2 Z^{ib}_{cd}v_bv^{\ast c}v^{\ast d}\equiv0 \text{ and }\frac{\partial V}{\partial \phi^{\ast j}}\rightarrow Y^a_j v_j+2 Z^{ab}_{jd}v_av_bv^{\ast d}.
\end{equation}
It is normally not possible to easily arrive at a closed form of the VEV, but it is implicitly defined as solution of the minimisation conditions. This allows to use the theorem about the derivatives of implicitly defined functions to arrive at a relation involving $\partial \hat v/\partial Y$ and $\partial \hat v/\partial Z$. Call a vector of the potential parameters $\vec z$. If $f(|v|,\hat v,\vec z)$ is the implicit function defining $|v|$ and $\hat v$, i.e.\ the minimisation condition, then an expression for $d \hat v/d \vec z$, which is identical to $\partial \hat v/\partial \vec z$, as $\hat v$ only depends on $\vec z$, can be obtained via the differential of $f$:
\begin{equation}
 df=\frac{\partial f}{\partial \vec z} d\vec z+\frac{\partial f}{\partial |v|} d|v|+\frac{\partial f}{\partial \hat v} d\hat v,
\end{equation}
from which follows
\begin{equation}
 \frac{\partial \hat v}{\partial \vec z}=-\left(\frac{\partial f}{\partial \hat v}\right)^{-1}\left(\frac{d f}{d \vec z}+\frac{\partial f}{\partial |v|}\frac{\partial |v|}{\partial \vec z}\right).
\end{equation}
While the above is interesting and the most direct line of thought, we did not manage to extract any practical criteria on the potential or its symmetries and do not follow it further in this work.

We had recently listed the minima obtained in the following in a short note, \cite{deMedeirosVarzielas:2017glw}, but without such a detailed discussion of the derivation as given here and without considering the effect of CP symmetries on the minima of the potentials of two triplets.

\subsection{Potentials of one triplet}
\subsubsection{$\bf\Delta(3n^2)$ and $\bf\Delta(6n^2)$, with $n>3$}
We start with a discussion of the simplest potential considered in this paper: that of one triplet of $\Delta(3n^2)$, which is identical to the potential of one triplet of $\Delta(6n^2)$.

In the notation from~\cite{Luhn:2007uq}, triplets of
$\Delta(3n^2)$ can be enumerated as ${\bf 3}_{(k,l)}$, where
$k,l=0,1,...,n-1$. The conjugate of ${\bf 3}_{(k,l)}$ is ${\bf 3}_{(-k,-l)}$.
For triplets in the same irrep.,
\be
{\bf 3}_{(k,l)} \otimes {\bf 3}_{(k,l)} ~=~ 
[{\bf 3}_{(2k,2l)} + {\bf 3}_{(-k,-l)}]_s + [{\bf 3}_{(-k,-l)}]_a \ .
\label{eq:kronecker3n2}
\ee
Here, $s$ and $a$ are the symmetric and antisymmetric
combinations.

This potential had not been analysed in \cite{Ivanov:2012fp, Ivanov:2014doa}, as the large discrete symmetry is so constraining that the renormalisable potential ends up having a continuous symmetry: apart from the $U(1)$ symmetry that make the potentials even in $\phi$, this potential has additional continuous symmetries, cf.\ Eq.~(\ref{1xD6n2_symmetries}). The potential is
\bea
V_{\Delta(6n^2)}(\varphi) = V_{\Delta(3n^2)}(\varphi)\equiv V_0 (\varphi)
\label{eq:potD3n2}
\eea
where, as it will appear in subsequent potentials, we define
\bea
V_0 (\varphi) \equiv
 - ~m^2_{\varphi}\sum_i   \varphi_i \varphi^{*i}
+ r \left( \sum_i   \varphi_i \varphi^{*i}  \right)^2
+ s \sum_i ( \varphi_i \varphi^{*i})^2.
\label{eq:potV0}
\eea
If each component of the triplet is an $SU(2)_L$ doublet,
\be
H = (h_{1\alpha},h_{2\beta},h_{3\gamma})\ ,
\ee
one additional
invariant appears, due to the two different ways to perform the $SU(2)_L$ contraction
on the discrete symmetry invariant $\left( \sum_i   \varphi_i \varphi^{*i}  \right)^2$,
when the $\varphi$ are replaced by Higgs doublets\footnote{Since the doublet 
  ${\bf{2}}$ of $SU(2)_L$ is a pseudoreal representation, it is also possible to
combine  $( h_{i \alpha} h_{j \beta} \epsilon^{\alpha\beta})( h^{*i\gamma}
h^{*j\delta} \epsilon_{\gamma\delta} )$  using the antisymmetric $\epsilon$
tensor. However, such a term is not independent of the two terms in
Eq.~\eqref{eq:doubletcontractions} as can be easily seen in an explicit
calculation or by noting that ${\bf{2\times 2=1+3}}$ which entails only two
independent  $SU(2)_L$ invariant quartic terms.},
\be
\sum_{i, j, \alpha, \beta} \left[
r_1 ( h_{i \alpha}  h^{*i\alpha})( h_{j \beta}  h^{*j\beta}) + r_2 ( h_{i
  \alpha}  h^{*i\beta})( h_{j \beta}  h^{*j\alpha}) \right]\ .
\label{eq:doubletcontractions}
\ee
Here 
we highlight the $SU(2)_L$ indices to clarify the distinct $SU(2)_L$ contractions.
Then we have
\bea
V_{\Delta(6n^2)}(H) = V_{\Delta(3n^2)}(H)\equiv V_0 (H)
\label{eq:potD3n2H}
\eea 
and we define $V_{0} (H)$ in analogy with Eq.~(\ref{eq:potV0}):
\begin{eqnarray}
 V_{0} ( H ) &=&
 - ~m^2_{h}\sum_{i, \alpha}   h_{i \alpha}  h^{*i\alpha}
+\sum_{i, j, \alpha, \beta} \left[ r_1 ( h_{i \alpha}  h^{*i\alpha})( h_{j \beta}  h^{*j \beta}) + r_2 ( h_{i \alpha}  h^{*i \beta})( h_{j \beta}  h^{*j\alpha}) \right]\notag \\
&&+ s \sum_{i, \alpha, \beta}  ( h_{i \alpha}  h^{*i\alpha})( h_{i \beta}
h^{*i\beta}).
\label{eq:potV0H}
\end{eqnarray}
The potential Eq.~(\ref{eq:potV0}) can be minimized analytically and one obtains for $m_\varphi \neq 0$ three classes of non-zero VEVs that are internally related by symmetry transformations and representatives of those classes are
\begin{equation}
v_1\cdot(1,0,0),\ v_2\cdot(1,1,0),\ v_3\cdot(1,1,1),
 \label{1xD3n2_alignments}
\end{equation}
where 
\begin{equation}
 v_1^2=\frac{m_\varphi^2}{2r+2s},\
 v_2^2=\frac{m_\varphi^2}{4r+2s},\
 v_3^2=\frac{m_\varphi^2}{6r+2s}.
\end{equation}
It can be tested that for all of the alignments listed previously, regions in the parameter space exist such that for potentials with parameters in that region, the corresponding alignment is the global minimum of the potential. 

The potential in Eq.~(\ref{eq:potV0}) decays into two parts, one, namely the two invariants with parameters $m_\varphi$ and $r$, invariant under all of $U(3)$, and a second part, consisting of the invariant with parameter $s$, invariant under $((U(1)\times U(1))\rtimes S_3)\times U(1)=:\Delta(6\infty^2)\times U(1)$, where the $U(1)^2$ factor in brackets arises from sending $n$ to infinity in $\Delta(6n^2)$, and the third $U(1)$ was imposed to keep the potential even:
\begin{equation}
 V_0=V_{U(3)}+V_{\Delta(6\infty^2)\times U(1)}.
\end{equation}

Occasionally, the label on a potential will refer to the imposed symmetry and occasionally, but only when talking about parts of potentials, to the full symmetry of that part of the potential. It is hoped that it is mostly clear from context which symmetry is meant.

The alignments of $V_{U(3)}$ all fall into one large orbit, represented e.g.\ by $(1,0,0)$, connected by arbitrary unitary transformations. The effect of the second, less symmetric, part of the potential is that this big orbit decays into several orbits in which now not yet phases but the direction of the VEV becomes physical.

To each of the alignments in Eq.~(\ref{1xD3n2_alignments}) quite a big orbit is attached, the members of which are related by the full symmetries of the potential, both such of flavour and CP type. The flavour symmetries of $V_0$ are generated by
\begin{equation}
 \begin{pmatrix}
  0&1&0\\0&0&1\\1&0&0
 \end{pmatrix},
 \begin{pmatrix}
  0&0&1\\0&1&0\\1&0&0
 \end{pmatrix},
  \begin{pmatrix}
  e^{i\alpha}&0&0\\0&e^{i\alpha}&0\\0&0&e^{i\alpha}
 \end{pmatrix}, 
  \begin{pmatrix}
  e^{i\beta}&0&0\\0&1&0\\0&0&e^{-i\beta}
 \end{pmatrix}, 
  \begin{pmatrix}
  1&0&0\\0&e^{i\gamma}&0\\0&0&e^{-i\gamma}
 \end{pmatrix},
 \label{1xD6n2_symmetries}
\end{equation}
where $\alpha,\beta,\gamma$ are arbitrary phases. Additionally, the potential is automatically invariant under canonical CP transformations, which we refer to as $CP_0$. Note that the alignments in Eq.~(\ref{1xD3n2_alignments}) all conserve canonical CP ($CP_0$). The orbits of alignments of this potential are
\begin{equation}
 \{\begin{pmatrix}
    e^{i\eta}\\0\\0
   \end{pmatrix},\begin{pmatrix}
   0\\e^{i\eta}\\0
   \end{pmatrix},\begin{pmatrix}
    0\\0\\e^{i\eta}
   \end{pmatrix}
\},\{\begin{pmatrix}
      e^{i\eta}\\e^{i\zeta}\\0
     \end{pmatrix},\text{permut.}
\},\{\begin{pmatrix}
      e^{i\eta}\\e^{i\zeta}\\e^{i\theta}
     \end{pmatrix},\text{permut.}
\}.
\label{1xD3n2_orbits}
\end{equation}
Now one could wonder into what kind of orbits the above alignments of Eq.~(\ref{1xD3n2_orbits}) are bundled by applying a smaller symmetry group than $\Delta(6n^2)\times U(1)$. In other words, one asks, which phases and permutations of the alignments of the one-triplet potential of $\Delta(6n^2)$, that were unphysical under the symmetries of Eq.\ (\ref{1xD6n2_symmetries}) would no longer be unphysical, assuming for a moment that they could actually be minima of a potential with a smaller symmetry. 

\subsubsection{$\bf S_4$}

The potential of one triplet of $S_4$ is 
\begin{equation}
 V_{S_4}(\varphi)=V_0(\varphi)+V_{S_4\times U(1)}(\varphi)
\end{equation}
with $V_0 (\varphi)$ from Eq.(\ref{eq:potV0}) and
\bea\label{eq:S4ini}
V_{S_4\times U(1)} (\varphi) &=& b \left(
\varphi_1 \varphi_1 \varphi^{*3} \varphi^{*3} +  \varphi_2 \varphi_2 \varphi^{*1} \varphi^{*1} + \varphi_3 \varphi_3 \varphi^{*2} \varphi^{*2} \right) \nonumber\\
&+& b \left(
\varphi^{*1} \varphi^{*1} \varphi_3 \varphi_3 + \varphi^{*2} \varphi^{*2} \varphi_1 \varphi_1 + \varphi^{*3} \varphi^{*3} \varphi_2 \varphi_2
 \right).
\eea
Note that $b$ is real.
The abbreviations cycl.\ to denote the cyclic
permutations, and h.c.\ to indicate the hermitian conjugate allow us to write the
potential of Eq.~\eqref{eq:S4ini} (and subsequent potentials) in a more compact way:
\begin{equation}
 V_{S_4\times U(1)}(\varphi)=b \left[ \left(\varphi_1 \varphi_1 \varphi^{*3} \varphi^{*3}
  +\text{cycl.}\right) + \text{h.c.} \right],
\label{eq:potS4}
\end{equation}
\begin{equation}
\begin{array}{rl}
\label{V24P}
 V_{S_4} (\varphi) = V_0 (\varphi) 
+b \left[ \left(\varphi_1 \varphi_1 \varphi^{*3} \varphi^{*3}
  +\text{cycl.}\right) + \text{h.c.} \right].
 \end{array}
\end{equation}
The $SU(2)_L$ doublet version is similar:
\begin{equation}
\label{V24H}
  V_{S_4} ( H ) =
 V_{0} (H)
+  \sum_{\alpha, \beta} b \left[\left(
h_{1 \alpha} h_{1 \beta}  h^{*3\alpha}  h^{*3\beta} + 
\text{cycl.}\right)  + \text{h.c.} \right].
\end{equation}
The flavour symmetries of the $S_4$ potentials, Eq.~(\ref{V24P}) and (\ref{V24H}) for $SU(2)$ singlets and doublets respectively, are generated by
\begin{equation}
 \begin{pmatrix}
  0&1&0\\0&0&1\\1&0&0
 \end{pmatrix},
  \begin{pmatrix}
  0&0&1\\0&1&0\\1&0&0
 \end{pmatrix},
  \begin{pmatrix}
  e^{i\alpha}&0&0\\0&e^{i\alpha}&0\\0&0&e^{i\alpha}
 \end{pmatrix}, 
  \begin{pmatrix}
  -1&0&0\\0&1&0\\0&0&-1
 \end{pmatrix}, 
  \begin{pmatrix}
  1&0&0\\0&-1&0\\0&0&-1
 \end{pmatrix}
 \label{1xS4_symmetries}
\end{equation}
and in addition, the potential is automatically invariant under $CP_0$.
Under these symmetries, the elements of the orbits of the potential of one triplet of $\Delta(6n^2)$, Eq.\ (\ref{1xD3n2_orbits}), fall into the following orbits:
\begin{equation}
 \begin{pmatrix}
  e^{i\eta}\\0\\0
 \end{pmatrix}\rightarrow
 \begin{pmatrix}
  1\\0\\0
 \end{pmatrix} ,
 \begin{pmatrix}
      e^{i\eta}\\e^{i\zeta}\\0
 \end{pmatrix}\rightarrow
 \begin{pmatrix}
  1\\e^{i\zeta'}\\0
 \end{pmatrix} \text{ with } \zeta'\in[0,\pi]
\end{equation}
and
\begin{equation}
 \begin{pmatrix}
      e^{i\eta}\\e^{i\zeta}\\e^{i\theta}
 \end{pmatrix}\rightarrow
 \begin{pmatrix}
  1\\e^{i\zeta'}\\e^{i\theta'}
 \end{pmatrix} \text{ with } \zeta''\in[0,\pi] \text{ and }\theta'\in [0,2\pi].
\end{equation}
What has happened so far is just that phases that were unphysical for $\Delta(6n^2)$ can now be physical. Minimizing the parts of the potential that depend on these potentially physical phases for the various orbits yields the following possible global minima of just the phase-dependent part, $V_{S_4\times U(1)}$,
\begin{equation}
 \begin{pmatrix}
  1\\0\\0
 \end{pmatrix},
 \begin{pmatrix}
  1\\1\\1
 \end{pmatrix},
 \begin{pmatrix}
  \pm1\\\omega\\\omega^2
 \end{pmatrix},
 \begin{pmatrix}
  1\\i\\0
 \end{pmatrix},
\end{equation}
which have been rephased into the form as they appear in \cite{Ivanov:2014doa}, and where we have defined
\begin{equation}
\omega \equiv e^{i 2 \pi/3}.
\end{equation}
The above alignments obtained in this way are, maybe surprisingly, in agreement with \cite{Ivanov:2014doa}. 

Two questions arise now: firstly, does e.g.\ the potential of one triplet of $S_4$ have global VEVs that don't arise from making unphysical phases in $V_{\Delta(6n^2)}$ physical in $V_{S_4\times U(1)}$, which would be in contradiction to \cite{Ivanov:2014doa}, at least for the potential of one triplet of $S_4$? And secondly, what is the situation when going from $\Delta(6n^2)$ to $\Delta(54)$, $\Delta(27)$ or $A_4$, or from $S_4$ to $A_4$, or more interestingly for the actual topic of this paper, from potentials of two copies of triplets without cross-terms to such with cross terms? These questions will be considered in the following, one after the other.

For the first question whether there can be ``rogue'' alignments, consider two VEV candidates $v_1, v_2$ (with no relation to $v_1$ and $v_2$ mentioned earlier) of, say, the potential of one triplet of $S_4$. If $v_1$ is a local minimum and for any other local minimum $v_2$ it holds that
\begin{equation}
 V_{S_4}(v_1)<V_{S_4}(v_2),
 \label{rogue_global_min}
\end{equation}
then $v_1$ is a global minimum of the potential. (If the potential is stable.) If now the situation could arise that $v_1$ is not a minimum of the partial potential $V_{\Delta(6\infty)\times U(1)}$, but instead is beaten by a VEV candidate $v_2$,
\begin{equation}
 V_{\Delta(6\infty)\times U(1)}(v_2)<V_{\Delta(6\infty)\times U(1)}(v_1),
\end{equation}
then one can still rearrange the summands of $V_{S_4}$, starting from Eq.~(\ref{rogue_global_min}), to arrive at
\begin{equation}
 V_{\Delta(6\infty)\times U(1)}(v_1)-V_{\Delta(6\infty)\times U(1)}(v_2)<V_{U(3)}(v_2)-V_{U(3)}(v_1)+V_{S_4\times U(1)}(v_2)-V_{S_4\times U(1)}(v_1).
\end{equation}
This means that by scaling the overall coefficient of $V_{\Delta(6\infty)\times U(1)}$ until above inequality is fulfilled but still non-zero, $v_1$ can be made a global minimum of the potential. As all parameters are non-zero, the symmetry of the potential is not enhanced in the corner of parameter space considered. This would indicate that there can be rogue global minima even when all parameters of the potential are non-zero. However, neither we or \cite{Ivanov:2014doa} found any and we conjecture that there is a symmetry reason behind this.

\subsubsection{$\bf A_4$ \label{1xA4_vev_section}}

The irreps. of $A_4$ are three one-dimensional representations and a real triplet. Two triplet products decompose as
\be
{\bf 3\otimes 3} ~=~ ({\bf 1}_0 + {\bf 1}_1 + {\bf 1}_2 + {\bf 3})_s + {\bf 3}_a\ . 
\label{eq:kronecker12}
\ee
where subscripts $s$ and
$a$, denote symmetric and antisymmetric combinations respectively. 
Throughout, we use the \cite{Ma:2001dn} basis for $A_4$ products.

We charge $\varphi$ under a $U(1)$ symmetry (or a discrete subgroup) forbidding terms such as
$\varphi_i \varphi_i$ or $\varphi_i \varphi_i \varphi_i$, for a more direct generalisation to cases where the triplets transform under the SM gauge group.
The renormalisable scalar potential invariant under $A_4$ and such a $U(1)$ is then:
\bea
\label{V12P}
V_{A_4} (\varphi) = V_0 (\varphi) +~ \left[ c \left(
\varphi_1 \varphi_1 \varphi^{*3} \varphi^{*3} + 
\text{cycl.} \right) ~+~ \text{h.c.}
 \right]
 \ ,
\eea
with $V_0 (\varphi)$ as defined in Eq.(\ref{eq:potV0}), and where $c$ is complex in general.

This potential is automatically invariant under the CP symmetry with:
\bea
X_{23} = \begin{pmatrix}
1 & 0 & 0 \\
0 & 0 & 1 \\
0 & 1 & 0 
\end{pmatrix},
\label{U23}
\eea
even for arbitrary complex coefficient $c$.

For $SU(2)_L$ doublets, the potential is similar
\begin{eqnarray}
\label{V12H}
 V_{A_4} ( H ) =
 V_{0} (H)
+ \sum_{\alpha, \beta} \left[ c \left(
h_{1 \alpha} h_{1 \beta}  h^{*3\alpha}  h^{*3\beta} + 
\text{cycl.} \right) + \text{h.c.} \right].
\end{eqnarray}
This potential is also invariant under a CP transformation that involves swapping the second and third component in flavour space while keeping $SU(2)_L$ contractions unchanged, i.e.\ $h_{2\alpha}\rightarrow h^{\ast 3\alpha}$ etc.:
\bea
X_{23}^H = \begin{pmatrix}
1 & 0 & 0 \\
0 & 0 & 1 \\
0 & 1 & 0 
\end{pmatrix} \otimes \delta^{\alpha}_{\beta}\ .
\label{U23H}
\eea
Therefore, CP is conserved automatically for this potential and all possible explicit CPIs necessarily vanish. 

One notes that the potential of one triplet of $A_4$, Eqs.~(\ref{V12P}) and (\ref{V12H}) for $SU(2)$ singlets and doublets respectively, is an extension of the potential of one triplet of $\Delta(3n^2)$ (in Eqs.~(\ref{eq:potD3n2},\ref{eq:potV0}) and Eq.~(\ref{eq:potD3n2H},\ref{eq:potV0H})) by a term that is invariant only under $A_4\times U(1)$: 
\begin{equation}
 V_{A_4}=V_0+ V_{A_4\times U(1)}.
\end{equation}
The full flavour-type symmetries of the full potential $V_{A_4}$ are generated by 
\begin{equation}
 \begin{pmatrix}0&1&0\\0&0&1\\1&0&0\end{pmatrix},
 \begin{pmatrix}e^{i\alpha}&0&0\\0&e^{i\alpha}&0\\0&0&e^{i\alpha}\end{pmatrix},
 \begin{pmatrix}-1&0&0\\0&1&0\\0&0&-1\end{pmatrix},
 \begin{pmatrix}1&0&0\\0&-1&0\\0&0&-1\end{pmatrix}.
 \label{1xA4_symmetries}
\end{equation}
and in addition the potential has the CP symmetry generated by $X_{23}$.
Under these symmetries, the elements of the orbits of the potential of one triplet of $\Delta(6n^2)$, Eq.\ (\ref{1xD3n2_orbits}), fall into the following orbits:
\begin{equation}
 \begin{pmatrix}
  e^{i\eta}\\0\\0
 \end{pmatrix}\rightarrow
 \begin{pmatrix}
  1\\0\\0
 \end{pmatrix} ,
 \begin{pmatrix}
      e^{i\eta}\\e^{i\zeta}\\0
 \end{pmatrix}\rightarrow
 \begin{pmatrix}
  1\\e^{i\zeta'}\\0
 \end{pmatrix} \text{ with } \zeta'\in[0,\pi]
 \label{1xA4_orbits_from_1xD3n2_orbits_1}
\end{equation}
and
\begin{equation}
 \begin{pmatrix}
      e^{i\eta}\\e^{i\zeta}\\e^{i\theta}
 \end{pmatrix}\rightarrow
 \begin{pmatrix}
  1\\e^{i\zeta'}\\e^{i\theta'}
 \end{pmatrix} \text{ with } \zeta''\in[0,\pi] \text{ and }\theta'\in [0,2\pi].
 \label{1xA4_orbits_from_1xD3n2_orbits_2}
\end{equation}

So far this has been straightforward and one has learned that out of the orbits of the potential invariant under $\Delta(6n^2)$, up to two phases can become physical, which now have to be determined by minimizing the parts of the potential that depend on them, which happens to be exactly $V_{A_4\times U(1)}$.
For example, plugging the alignment $(1,e^{i\zeta'},0)$ into $V_{A_4\times U(1)}$, one obtains
\begin{equation}
 V_{A_4\times U(1)}[(1,e^{i\zeta'},0)]=c e^{2i\zeta'}+c^\ast e^{-2i\zeta'}, 
 \label{A4_alpha_condition}
\end{equation}
which when minimizing this in $\zeta'$ of course yields $\zeta'=-\text{Arg}(c)/2 \text{ mod }\pi$.

For the alignment candidate $(  1,e^{i\zeta''},e^{i\theta'})$ one obtains after a only slightly longer calculation for the phases which minimize $V_{A_4\times U(1)}$ and after eliminating phase constellations that are related by symmetries of the potential, Eq.~(\ref{1xA4_symmetries}),
\begin{equation}
 (\zeta'',\theta')=(0,0),(\pi/3,5 \pi/3),(2 \pi/3,4 \pi/3),
\end{equation}
which correspond to alignments of the type $(1,1,1)$ and both of $(\pm 1,\omega,\omega^2)$, such that with Eq.~(\ref{1xA4_orbits_from_1xD3n2_orbits_1}) the list of possible global alignments from \cite{Toorop:2010ex, Toorop:2010kt} and \cite{Ivanov:2014doa} for one triplet of $A_4$ is completed. All of the alignments found above, in summary,
\begin{equation}
 \begin{pmatrix}
  1\\0\\0
 \end{pmatrix},
 \begin{pmatrix}
  1\\1\\1
 \end{pmatrix},
 \begin{pmatrix}
  \pm1\\\omega\\\omega^2
 \end{pmatrix},
 \begin{pmatrix}
  1\\e^{i\zeta'}\\0
 \end{pmatrix}
 \label{1xA4_VEVs}
\end{equation}
have the same energy for $V_0$  and are just further differentiated by $V_{A_4\times U(1)}$, as is well known in the literature.

There are two ways in which one can reach the symmetries of the $A_4$ potential from the symmetries of the $\Delta(6n^2)$ potential, namely directly and via the symmetries of the $S_4$ potential.

The potential of one triplet of $S_4$ is actually contained in the potential of one triplet of $A_4$:
\begin{equation}
 V_{A_4}=V_0+V_{S_4\times U(1)}+V'_{A_4\times U(1)}
\end{equation}
with $V_{S_4 \times U(1)}$ from Eq.~(\ref{eq:potS4}) (with $b$ real and the $h.c.$ inside the bracket) and
\begin{equation}
 V'_{A_4\times U(1)}=(c-b)\left(
\varphi_{1} \varphi_{1} \varphi^{\ast3} \varphi^{\ast3} + \varphi_{2} \varphi_{2} \varphi^{\ast1} \varphi^{\ast1} + \varphi_{3} \varphi_{3} \varphi^{\ast2} \varphi^{\ast2}\right)
 +h.c. 
\end{equation}
Note that while $b$ is real, $c$ is complex and in this expression the $h.c.$ is outside of the bracket.
Following the same path, namely writing out the orbits of the VEVs of the $S_4$ potential and reducing them by the full symmetries of the $A_4$ potential, one recovers the first three VEVs in Eq.~(\ref{1xA4_VEVs}). However, there is no way of obtaining $(1,e^{i\zeta'},0)^T$ from the $S_4\times U(1)$ orbit of $(1,i,0)^T$: The full symmetry of the potential invariant under $S_4$ is such that the orbit of $(1,i,0)^T$ does not contain $(1,e^{i\zeta'},0)$. Interestingly, this $A_4$ VEV is not geometric.

\subsubsection{$\bf \Delta(27)$ and $\bf \Delta(54)$}

As in the $A_4$ case, we start with a single triplet of SM singlets and later consider up to two $\Delta(27)$ triplets of $SU(2)_L$ doublets.

$\Delta(27)$ has irreps. ${\bf 3}$, its
conjugate $\bar{\bf 3}$, and nine one-dimensional irreps.
The product of triplets decomposes as
\be
{\bf 3 \otimes 3} ~=~ ({\bf \ol 3 + \ol 3})_s ~+~ {\bf \ol 3}_a \ ,
\label{eq:Kronecker27} 
\ee
$s$ and $a$ are the symmetric and antisymmetric
combinations, respectively.

In the basis used in \cite{deMedeirosVarzielas:2006fc,Ma:2006ip}The renormalisable scalar
potential or one triplet is , 
\bea \label{V27P}
V_{\Delta(27)} (\varphi) &=&
V_0(\varphi)
+~ \left[d \left(
\varphi_1 \varphi_1 \varphi^{*2} \varphi^{*3} + 
\text{cycl.} \right) +\text{h.c.}\right]
 \ .
\eea
Apart from $d\in \mathbb C$, the coefficients are real. Note that this potential has no automatic CP symmetry. In the literature, for the potential Eq.~(\ref{V27P}), a variety of CP symmetries have been discussed, starting with 12 CP transformations that had been found to be consistent with $\Delta(27)$ in \cite{Nishi:2013jqa}. The enumeration of the $X$ matrices here follows that paper. The full discrete flavour symmetry of the potential invariant under $\Delta(27)$ is $\Delta(54)$ and in this case, pairs of $X$ matrices become related under the enlarged symmetry. In the following, the $X$ matrices are listed by the relation they enforce on the parameters of the potential. After that, basis transformations that relate these CP symmetries, and their role are discussed.

\label{Nishi_CP_and_basis_trafos}
\begin{itemize}
 \item $\text{Arg}(d)=0$
 \begin{equation}
  X_0=\begin{pmatrix}1&0&0\\0&1&0\\0&0&1\end{pmatrix}~\text{ or
 }~~X_1=\begin{pmatrix}1&0&0\\0&0&1\\0&1&0\end{pmatrix}
 \end{equation}
\item $\text{Arg}(d)=4\pi/3$
\begin{equation}
 X_2=\begin{pmatrix}1&0&0\\0&1&0\\0&0&\omega\end{pmatrix}\text{ or }~~X_8=\begin{pmatrix}\omega&0&0\\0&0&1\\0&1&0\end{pmatrix}
\end{equation}
\item $\text{Arg}(d)=2\pi/3$
\begin{equation}
 X_3=X_2^*=\begin{pmatrix}1&0&0\\0&1&0\\0&0&\omega^2\end{pmatrix}\text{ or }~~X_9=X_8^*=\begin{pmatrix}\omega^2&0&0\\0&0&1\\0&1&0\end{pmatrix}\label{X3}
\end{equation}
\item $s=(d+d^*)=2 \mathrm{Re} (d)$
\begin{equation}
 \label{X4} X_4=\frac{1}{\sqrt{3}}\begin{pmatrix}1&1&1\\1&\omega&\omega^2\\1&\omega^2&\omega\end{pmatrix} \text{ or }~
 X_5=X_4 X_1=X_4^\ast=\frac{1}{\sqrt{3}}
\left(
\begin{array}{ccc}
 1 & 1 & 1 \\
 1 & \omega^2 & \omega \\
 1 & \omega & \omega^2 \\
\end{array}
\right)
\end{equation}
\item $s=-\mathrm{Re} (d) - \sqrt{3} \mathrm{Im} (d)$
\begin{equation}
 X_6=\frac{-i}{\sqrt{3}}
\left(
\begin{array}{ccc}
 1 & \omega & \omega \\
 \omega & \omega & 1 \\
 \omega & 1 & \omega \\
\end{array}
\right)
\text{ or }~X_{10}=X_6 X_1=\frac{-i}{\sqrt{3}}\left(
\begin{array}{ccc}
 1 & \omega & \omega \\
 \omega & 1 & \omega \\
 \omega & \omega & 1 \\
\end{array}\right)
\end{equation}
\item $s=-\mathrm{Re} (d) + \sqrt{3} \mathrm{Im} (d)$
\begin{equation}
 X_7=X_6^\ast=\frac{i}{\sqrt{3}}\left(
\begin{array}{ccc}
 1 & \omega^2 & \omega^2 \\
 \omega^2 & \omega^2 & 1 \\
 \omega^2 & 1 & \omega^2 \\
\end{array}
\right)\text{ or }~X_{11}=X_7 X_1=\frac{i}{\sqrt{3}}\left(
\begin{array}{ccc}
 1 & \omega^2 & \omega^2 \\
 \omega^2 & 1 & \omega^2 \\
 \omega^2 & \omega^2 & 1 \\
\end{array}
\right)
\end{equation}

\end{itemize}

As all of the above $X$ matrices are symmetric, there are basis transformations $U_i$ such that in the new basis the matrix $X_i$ becomes the identity, and for $X_2$ and $X_3$, these also leave the potential form invariant, as was realised in \cite{Fallbacher:2015rea},
\begin{equation}
  U_2=\begin{pmatrix}1&0&0\\0&1&0\\0&0&\omega^2\end{pmatrix} \text{ and } U_3=\begin{pmatrix}1&0&0\\0&1&0\\0&0&\omega\end{pmatrix}.
\end{equation}

There are of course also basis transformations from $X_4,X_5,X_6$ to $X_0$, but these do not leave the potential form-invariant. The basis transformation e.g.\ from $X_4$ to $X_0$ is
\begin{align}
  U_4= \left(
\begin{array}{ccc}
 0 & 0 & 1 \\
 -\frac{1}{\sqrt{2}} & \frac{1}{\sqrt{2}} & 0 \\
 \frac{1}{\sqrt{2}} & \frac{1}{\sqrt{2}} & 0 \\
\end{array}
\right)
\left(
\begin{array}{ccc}
 1 & 0 & 0 \\
 0 & -\frac{1+\sqrt{3}}{\sqrt{6+2\sqrt{3}}} & \frac{1}{\sqrt{3+\sqrt{3}}} \\
 0 & \frac{\sqrt{3}-1}{\sqrt{6-2 \sqrt{3}}} & \frac{1}{\sqrt{3-\sqrt{3}}} \\
\end{array}
\right)
\left(
\begin{array}{ccc}
 -(-1)^{3/4} & 0 & 0 \\
 0 & -i & 0 \\
 0 & 0 & 1 \\
\end{array}
\right).
\end{align}
For $SU(2)_L$ doublets, the
potential is
\bea \label{V27H}
V_{\Delta(27)} (H) = V_0(H) 
~+~ \sum_{\alpha, \beta} \left[d \left(
h_{1 \alpha} h_{1 \beta}  h^{*2 \alpha}  h^{*3 \beta} + 
\text{cycl.} \right) +
\text{h.c.}\right]
.
\eea
For one triplet, the corresponding potentials invariant under are $\Delta(54)$ identical to those invariant under $\Delta(27)$~\cite{deMedeirosVarzielas:2011zw}
\bea
V_{\Delta(54)} (\varphi)= V_{\Delta(27)}(\varphi) \ , \label{V54P} \\
V_{\Delta(54)} (H)= V_{\Delta(27)}(H). \label{V54H}
\eea
The potential Eq.~(\ref{V27H}) has been analysed in \cite{Ivanov:2014doa} and is the 3HDM potential with the largest of the purely discrete symmetries (apart for the $U(1)$ that arises from the potential being even) for which CP may be spontaneously violated. 
In addition, this was the case where so-called geometric violation was first discovered, see also section \ref{sec:S_D27}.
   
 For generic values of the potential parameters, the potential violates CP explicitly. Additional CP symmetries may be imposed that force the potential to be explicitly CP-conserving. In \cite{Ivanov:2014doa} the two types of CP symmetry that are normally considered consistent with the flavour-type symmetry of the potential are analysed. In \cite{Nishi:2013jqa}, 12 CP symmetries are listed as being consistent with $\Delta(27)$, but if considered in the $\Delta(54)$ context this number reduces to 6. For example, $CP_0$ and the $CP_{23}$ (the CP associated with $X_{23}$) become related by $\Delta(54)$. In this case, of the 6 remaining, 3 restrict the phase of parameter $d$ in the potential, the remaining 3 additionally enforces an additional relation on the parameters, such as $2s=(d+d^*)$ if CP with $X$ matrix
\begin{equation}
X_4=\frac{1}{\sqrt{3}}\begin{pmatrix}1&1&1\\1&\omega&\omega^2\\1&\omega^2&\omega\end{pmatrix},
\label{X4_delta27}
\end{equation}
is imposed. 
This second CP-type transformation (like $X_4$) enlarges the flavour-type symmetry of the potential to $\Sigma(36)$.
It should be noted that the three CP symmetries that restrict the phase of $d$ are related by basis-transformations that leave the potential form-invariant, cf.\ \cite{Fallbacher:2015rea}. In addition, we found that the three CP symmetries that enforce $2s=d+d^\ast$ are also related with each other via such basis transformations, the simplest ones being for connecting $X_4$ with $X_{10}$ and $X_{11}$, respectively,
\begin{equation}
 U_{10}=\sqrt{-i}\begin{pmatrix}
  1&0&0\\0&\omega&0\\0&0&\omega
 \end{pmatrix}
~\text{and}~
 U_{11}=\sqrt{-i}\begin{pmatrix}
  1&0&0\\0&\omega^2&0\\0&0&\omega^2
 \end{pmatrix}.
\end{equation}
Furthermore, as $X_4$ of Eq.~(\ref{X4_delta27}) is also symmetric, in principle a basis transformation exists, such that in the new basis $X_4$ would be diagonal, however this basis transformation does not leave the potential form-invariant.

The full flavour-type symmetries of an even potential of one triplet of $\Delta(27)$ or one triplet of $\Delta(54)$, both produce identical renormalisable potentials, are generated by
\begin{equation}
 \begin{pmatrix}
  0&1&0\\0&0&1\\1&0&0
 \end{pmatrix},
  \begin{pmatrix}
  0&0&1\\0&1&0\\1&0&0
 \end{pmatrix},
  \begin{pmatrix}
  e^{i\alpha}&0&0\\0&e^{i\alpha}&0\\0&0&e^{i\alpha}
 \end{pmatrix}, 
  \begin{pmatrix}
  e^{i2\pi/3}&0&0\\0&1&0\\0&0&e^{-i2\pi/3}
 \end{pmatrix}, 
  \begin{pmatrix}
  1&0&0\\0&e^{i2\pi/3}&0\\0&0&e^{-i2\pi/3}
 \end{pmatrix}.
 \label{1xD54_symmetries}
\end{equation}
Using these, the orbits of one triplet of $\Delta(6n^2)$, Eq.~(\ref{1xD3n2_orbits}) become 
\begin{equation}
 (1,0,0),~(1,e^{i\beta'},0),~(1,e^{i\beta''},e^{i\gamma'})
\end{equation}
with $\beta',\beta'',\gamma'\in[0,2\pi/3]$. The CP symmetries, whether $CP_0$ or the CP symmetry associated with $X_4$ do at this point not further constrain the potentially physical phases. The phases appearing here can now be physical within $\Delta(54)$ and need to be determined by minimizing the parts of the potential that depend on them. Curiously, the phase dependent part of  $V_{\Delta(54)}$,
\begin{equation}
 V_{\Delta(54)\times U(1)}= \left[d \left(
\varphi_1 \varphi_1 \varphi^{*2} \varphi^{*3} + 
\text{cycl.} \right) +\text{h.c.}\right]
\end{equation}
yields simply zero for the alignment $(1,e^{i\beta'},0)$, which already means that the phase $\beta'$ remains unphysical. $(1,1,0)$ could still be a local minimum of the potential, as it already is a possible global minimum of $V_0$. When minimizing $V_{\Delta(54)\times U(1)}$ with $(1,e^{i\beta''},e^{i\gamma'})$, one obtains the usual alignments of $(1,1,1),~(1,1,\omega),(1,\omega,\omega)$ (or for the last one equivalently $(1,1,\omega^2)$). Evaluating the potential at those alignments explains the fate of the $(1,1,0)$ alignment as a global minimum: for any value of $\text{Arg}(d)$, for one of $(1,1,1)$, $(1,1,\omega)$, $(1,1,\omega^2)$, $V_{\Delta(54)\times U(1)}$ has a negative value. However, as this scales with the absolute size of $d$ as well, $(1,1,0)$ could still be quite a metastable minimum. In any case, again, taking VEV candidates just from the VEV orbits of the $\Delta(6n^2)$ potential produces the full list of global minima that were obtained (painstakingly) by \cite{Ivanov:2014doa}, in summary:
\begin{equation}
 (1,0,0),(1,1,1),(1,1,\omega),(1,1,\omega^2).
 \label{1xD54_VEVs}
\end{equation}
To also repeat the effect of CP transformations from \cite{Ivanov:2014doa}:
With $CP_0$, the last two VEVs become related by symmetry and no longer represent different breaking patterns. With the type of CP that extends the flavour symmetry to $\Sigma(36)$, with matrix $X_4$, Eq.~(\ref{X4_delta27}), again the last two VEVs in Eq.~(\ref{1xD54_VEVs}) become part of the same orbit. In addition also the first two VEVs in Eq.~(\ref{1xD54_VEVs}) become part of the same orbit (separate from the last two VEVs).

\subsection{VEVs of potentials of two triplets}

Typically, realistic models of flavour require more than just one triplet 
flavon. We therefore consider potentials involving two physically different flavon
fields $\varphi$ and $\varphi'$ which both transform under a triplet representation of the symmetry groups we consider.
Similarly we also consider 6HDMs with the $SU(2)_L$ doublets transforming as two triplets of the discrete symmetries.

Potentials of two triplets consist of three parts: two sets of terms that each only couple components of one triplet to each other on one hand, and cross terms that couple components of different triplets to each other,
\begin{equation}
 V(\varphi,\varphi')=V(\varphi)+V'(\varphi')+V_c(\varphi,\varphi').
 \label{eq:cross}
\end{equation}
If the two triplets transform identically under the symmetry, then $V(\varphi)$ and $V'(\varphi')$ will be functionally identical. 

The orbits of minima of the potentials of single triplets are known completely for the flavour symmetries in consideration. 
Similarly to potentials of one triplet, one can now analyse which degrees of freedom of VEVs that are unphysical by the symmetry of the two single-triplet potentials $V(\varphi)+V'(\varphi')$ can become physical by reducing the symmetry either by introducing cross-terms between triplets, $V_c(\varphi,\varphi')$, or by considering a whole potential invariant only under a smaller symmetry group.

As before, only the alignments of the VEVs  are shown and for two triplets, this does not mean that the VEVs $v_1,v_2$ of both triplets have to have the same length. Instead, arbitrary lengths of both triplets are allowed and possible and generally $|v_1|\neq|v_2|$. 

For use in the following subsections, we define \bea
V_1 (\varphi,\varphi') &=&
+~ \tilde r_1 \left( \sum_i \varphi_i \varphi^{*i} \right)
\left( \sum_j \varphi'_j \varphi'^{*j} \right) 
+ \tilde r_2\left( \sum_i \varphi_i \varphi'^{*i} \right)
\left( \sum_j \varphi'_j \varphi^{*j} \right) \notag \\[2mm]
&& +~ \tilde s_1\sum_i \left(\varphi_i \varphi^{*i} \varphi'_i \varphi'^{*i}
\right) \notag \\[2mm]
&& +~ \tilde s_2 \left(
\varphi_1 \varphi^{*1} \varphi'_2 \varphi'^{*2} + 
\varphi_2 \varphi^{*2} \varphi'_3 \varphi'^{*3} + 
\varphi_3 \varphi^{*3} \varphi'_1 \varphi'^{*1} 
\right)  \notag \\[2mm]
&& +~ i \, \tilde s_3 
\Big[
(\varphi_1 \varphi'^{*1} \varphi'_2 \varphi^{*2} + \text{cycl.}
) 
- 
( \varphi^{*1}\varphi'_1  \varphi'^{*2} \varphi_2 +\text{cycl.}
)
\Big].
\label{eq:potV1}
\eea
Note that in this definition, the term multiplied by $\tilde{r}_1$ contains
the term multiplied by $\tilde{s}_2$ as well as the term obtained from the
latter by interchanging $\varphi$ with $\varphi'$,
\be
\left(
\varphi'_1 \varphi'^{*1} \varphi_2 \varphi^{*2} + 
\varphi'_2 \varphi'^{*2} \varphi_3 \varphi^{*3} + 
\varphi'_3 \varphi'^{*3} \varphi_1 \varphi^{*1} 
\right) ,
\ee
which is not included separately in $\tilde{s}_2$.

Earlier, when considering a potential of one triplet of $SU(2)_L$ doublets, the only difference was that the term with coefficient $r$ split into two different invariants corresponding to two different possible $SU(2)_L$ contractions, cf.\ Eq.~(\ref{eq:doubletcontractions}). Similarly, the potential of two triplets of SM doublets,
\be
H = (h_{1\alpha},h_{2\beta},h_{3\gamma}) \,, \quad H' =
(h'_{1\alpha},h'_{2\beta},h'_{3\gamma})\ ,
\ee
can be obtained from the corresponding potential of singlets,
Eq.~(\ref{eq:potV1}). In the first two parts of the potential,
  $V_0(\varphi)$ and $V_0(\varphi')$, as earlier, there are two different ways
  of $SU(2)_L$-contracting the invariants with coefficients $r$ and $r'$. In
  the potential $V_1(\varphi,\varphi')$, there are two possible ways of $SU(2)_L$-contracting for each invariant and therefore this part of the potential becomes in its $SU(2)_L$ doublet version
\begin{align}
 V_1(H,H')&=
\sum_{i,j,\alpha,\beta} \left[
\tilde{r}_{11}h_{i\alpha}h^{\ast i \alpha}h_{j \beta}'h'^{\ast j
  \beta}+\tilde{r}_{12}h_{i \alpha}h'^{\ast j \alpha} h'_{j \beta}h^{\ast i
  \beta}
\right] \notag \\[2mm]
 &+\sum_{i,j,\alpha,\beta} \left[
\tilde{r}_{21}h_{i\alpha}h'^{\ast i \alpha}h'_{j \beta}h^{\ast j
  \beta}+\tilde{r}_{22}h_{i \alpha}h^{\ast j \alpha}h'_{j \beta}h'^{\ast i
  \beta}
\right]\notag \\[2mm]
 &+\sum_{i,\alpha,\beta} \left[
\tilde{s}_{11}h_{i \alpha}h^{\ast i \alpha}h'_{i \beta}h'^{\ast i
  \beta}+\tilde{s}_{12}h_{i \alpha}h'^{\ast i \alpha}h'_{i \beta}h^{\ast i
  \beta}
\right]\notag \\[2mm]
 &+\sum_{\alpha,\beta} \left[
\tilde{s}_{21}(h_{1\alpha}h^{\ast1\alpha}h'_{2\beta}h'^{\ast2\beta}+\text{cycl.})+\tilde{s}_{22}(h_{1\alpha}h'^{\ast
  2 \alpha}h'_{2\beta}h^{\ast 1\beta}+\text{cycl.})
\right]\notag \\[2mm]
 &+i\tilde{s}_{31}\sum_{\alpha,\beta} 
[(h_{1\alpha}h'^{\ast1\alpha}h'_{2\beta}h^{\ast2\beta}+\text{cycl.}) - (h^{\ast1\alpha}h'_{1\alpha}h'^{\ast2\beta}h_{2\beta}+\text{cycl.})]\notag \\[2mm]
 &+i\tilde{s}_{32}\sum_{\alpha,\beta} [(h_{1\alpha}h^{\ast 2 \alpha}h'_{2\beta}h'^{\ast1\beta}+\text{cycl.}) - (h^{\ast1\alpha}h_{2 \alpha}h'^{\ast2\beta}h'_{1\beta}+\text{cycl.})].
\label{eq:potV1H}
 \end{align}
We define also
\bea\label{eq:potV2}
V_2 (\varphi,\varphi') &=&
 \tilde r_1 \left( \sum_i \varphi_i \varphi^{*i} \right)
\left( \sum_j \varphi'_j \varphi'^{*j} \right) 
+ \tilde r_2\left( \sum_i \varphi_i \varphi'^{*i} \right)
\left( \sum_j \varphi'_j \varphi^{*j} \right) \notag \\[2mm]
&&
 +~ \tilde s_1\sum_i \left(\varphi_i \varphi^{*i} \varphi'_i \varphi'^{*i}
\right), 
\eea
and the $SU(2)_L$ doublet version
\bea
 V_2(H,H')&=&\sum_{i,j,\alpha,\beta}\left[
\tilde{r}_{11}h_{i\alpha}h^{\ast i \alpha}h_{j \beta}'h'^{\ast j
  \beta}+\tilde{r}_{12}h_{i \alpha}h'^{\ast j \alpha} h'_{j \beta}h^{\ast i
  \beta} \right]\notag \\
& &+\sum_{i,j,\alpha,\beta}\left[
\tilde{r}_{21}h_{i\alpha}h'^{\ast i \alpha}h'_{j \beta}h^{\ast j \beta}+\tilde{r}_{22}h_{i \alpha}h^{\ast j \alpha}h'_{j \beta}h'^{\ast i \beta}\right]\notag \\
 &&+\sum_{i,\alpha,\beta}\left[
\tilde{s}_{11}h_{i \alpha}h^{\ast i \alpha}h'_{i \beta}h'^{\ast i \beta}+\tilde{s}_{12}h_{i \alpha}h'^{\ast i \alpha}h'_{i \beta}h^{\ast i \beta}\right].
\label{eq:potV2H}
 \eea
We note that one can obtain $V_2$ from $V_1$ by imposing
\bea
\tilde{s}_2 = \tilde{s}_3 = 0\ ,
\label{3n2_to_6n2}
\eea
for $SU(2)_L$ singlets and, for $SU(2)_L$ doublets,
$s_{22}=\tilde s_{31}=\tilde s_{32}=0$. 

As previously for one-triplet-potentials, minima of several doublets that conserve $U(1)_{em}$ can identified with the minima of the same number of singlets.

By restricting the potential with $U(1)$ symmetries for each of the scalar fields, the mixed terms appearing are limited to the form  $\varphi\,\varphi' \, 
\varphi^* \, \varphi'^*$. For $SU(2)_L$ doublets, it is even sufficient to impose e.g.\ a $Z_3$ symmetry with non-trivial charge for only one of the two triplets of Higgs doublets to distinguish them. 
\subsubsection{$\bf \Delta(6n^2)$ with $n>3$}

The potentials of two triplets under $\Delta(6n^2)$ of singlets and doublets have the form
\bea
V_{\Delta(6n^2)} (\varphi,\varphi')&=& V_0(\varphi)  + V'_0(\varphi')  + V_2(\varphi,\varphi')  \ , \label{V6n2PP} \\[2mm]
V_{\Delta(6n^2)} (H,H')&=& V_0(H) + V'_0(H') + V_2(H,H')\ \label{V6n2HH},
\eea
where $V_0' (\varphi')$ has the same functional form as $V_0 (\varphi)$  with
different coefficients $m'_{\varphi'}$, $r'$, $s'$ and depends on $\varphi'$.
$V_0(\varphi)$ and $V_0(H)$ had been defined in Eqs.~(\ref{eq:potV0}) and (\ref{eq:potV0H}).
These $\Delta(6n^2)$ potentials (with $n>3$) conserve CP
explicitly. As the parameters are all real, one of the CP symmetries is $CP_0$.

The orbits of minima of a potential of one triplet of $\Delta(6n^2)$ are (cf.\ Eq.~(\ref{1xD3n2_orbits}))
\begin{equation}
 \{\begin{pmatrix}
    e^{i\eta}\\0\\0
   \end{pmatrix},\begin{pmatrix}
   0\\e^{i\eta}\\0
   \end{pmatrix},\begin{pmatrix}
    0\\0\\e^{i\eta}
   \end{pmatrix}
\},\{\begin{pmatrix}
      e^{i\eta}\\e^{i\zeta}\\0
     \end{pmatrix},\text{permut.}
\},\{\begin{pmatrix}
      e^{i\eta}\\e^{i\zeta}\\e^{i\theta}
     \end{pmatrix},\text{permut.}
\}.
\label{1xD6n2_orbits}
\end{equation}
Without cross terms $V_2(\varphi,\varphi')$, both triplets can transform independently under $\Delta(6n^2)$ and thus have independent orbits.
Thus, every member of each orbit may now be combined with every other member of each orbit, but only the symmetries of the full potential of two triplets including cross terms can be used to eliminate unphysical phases. 
The symmetries of the full potential are generated by the simultaneous transformations of both triplets under $\Delta(6\infty^2)$ and in addition separate $U(1)$ phases acting on each triplet,
\begin{equation}
\begin{pmatrix}
e^{i\alpha}&&\\&e^{i\alpha}&\\&&e^{i\alpha} 
\end{pmatrix}\oplus
\begin{pmatrix}
1&&\\&1&\\&&1 
\end{pmatrix}\text{, and }
\begin{pmatrix}
1&&\\&1&\\&&1 
\end{pmatrix}\oplus
\begin{pmatrix}
e^{i\alpha'}&&\\&e^{i\alpha'}&\\&&e^{i\alpha'} 
\label{two_triplet_phases}
\end{pmatrix},
\end{equation}
as well as an overall canonical CP transformation ($CP_0$). With these symmetries one obtains the following reduction  of  combinations of orbits of VEVs of a potential of one triplet of  $\Delta(6n^2)$:
\begin{align}
 (e^{i\eta},0,0),(e^{i\eta'},0,0)&\rightarrow (1,0,0),(1,0,0) \label{2xDelta6n2vevs_start}\\
 (e^{i\eta},0,0),(0,e^{i\eta'},0)&\rightarrow (1,0,0),(0,1,0)\\
 (e^{i\eta},0,0),(e^{i\eta'},e^{i\zeta'},0)&\rightarrow (1,0,0),(1,1,0)\\
 (e^{i\eta},0,0),(0,e^{i\eta'},e^{i\zeta'})&\rightarrow (1,0,0),(0,1,1)\\
 (e^{i\eta},0,0),(e^{i\eta'},e^{i\zeta'},e^{i\theta'})&\rightarrow (1,0,0),(1,1,1)\\
  (e^{i\eta},e^{i\zeta},0),(0,e^{i\zeta'},e^{i\theta'})&\rightarrow (1,1,0),(0,1,1)\\
 (e^{i\eta},e^{i\zeta},0),(e^{i\eta'},e^{i\zeta'},0)&\rightarrow (1,1,0),(1,e^{i\zeta'},0)\\
 (e^{i\eta},e^{i\zeta},0),(e^{i\eta'},e^{i\zeta'},e^{i\theta'})&\rightarrow (1,1,0),(1,e^{i\zeta'},1)\\
 (e^{i\eta},e^{i\zeta},e^{i\theta}),(e^{i\eta'},e^{i\zeta'},e^{i\theta'})&\rightarrow (1,1,1),(1,e^{i\zeta'},e^{i\theta'}). \label{2xDelta6n2vevs_end}
\end{align}
Note that at this stage orbit pairs that arise from interchanging first and second triplet are redundant, which is why e.g.\ $(1,1,0),(1,0,1)$ is not listed.
The remaining phases are determined by minimizing the parts of the potential that depend on them and one obtains for both of $(1,1,0),(1,e^{i\zeta'},0)$ and $(1,1,0),(1,e^{i\zeta'},1)$ that $\zeta'=0$ for $r_2'>0$ and $\zeta'=\pi$ for $r_2<0$, leading to the following alignments:
\begin{equation}
 (1,1,0),(1,\pm1,0) \text{ and } (1,1,0),(1,\pm1,1)
\end{equation}
where different sign choices correspond to different orbits.
For $(1,1,1),(1,e^{i\zeta'},e^{i\theta'})$ minimizing the part of the potential depending on $\zeta'$ and $\theta'$ produces for $r_2<1$ the orbit
\begin{equation}
 (1,1,1),(1,1,1)
\end{equation}
and for $r_2>0$ the two orbits
\begin{equation}
 (1,1,1),(1,\omega,\omega^2)\text{ and }(1,1,1),(1,\omega^2,\omega).
\end{equation}
To summarize all possible pairs:
\begin{align}
 (1,0,0),(1,0,0)\label{2xDelta6n2vevsummary}\\
 (1,0,0),(0,1,0)\\
 (1,0,0),(1,1,0)\\
 (1,0,0),(0,1,1)\\
 (1,0,0),(1,1,1)\\
 (1,1,0),(0,1,1)\\
   (1,1,0),(1,\pm1,0)\\
   (1,1,0),(1,\pm1,1)\\
    (1,1,1),(1,1,1)\\
    (1,1,1),(1,\omega,\omega^2)\\
    (1,1,1),(1,\omega^2,\omega)
\end{align}

\subsubsection{$\bf \Delta(3n^2)$ with $n>3$}
Next, consider the potential of two triplets of $\Delta(3n^2)$ with $n>3$,
\bea
V_{\Delta(3n^2)} (\varphi,\varphi') &=&
V_0 (\varphi) + V_0'(\varphi') + V_1 (\varphi, \varphi'),
\label{V3n2PP}
\eea
with $V_0(\varphi)$ and $V_1(\varphi,\varphi')$ defined in Eqs.~(\ref{eq:potV0}) and (\ref{eq:potV1}) for singlets and 
\bea
 V_{\Delta(3n^2)} (H,H') &= V_0(H) + V_0'(H')+ V_1(H,H')\ \label{V3n2HH}
\eea
for doublets, where $V_0(H)$ and $V_1(H,H')$ were defined in Eqs.~(\ref{eq:potV0H}) and (\ref{eq:potV1H}).

The orbits of VEVs of the corresponding potential of one triplet are identical to that of one triplet of $\Delta(6n^2)$, cf.\ Eq.~(\ref{1xD6n2_orbits}).
The difference to the previous potential lies in the fact that the full symmetries of $V_{\Delta(3n^2)}$ only allow for cyclic permutations, i.e.\ only
\begin{equation}
 \begin{pmatrix}
  &1&\\&&1\\1&&
 \end{pmatrix}\oplus
 \begin{pmatrix}
  &1&\\&&1\\1&&
 \end{pmatrix}
\end{equation}
with otherwise identical phase symmetries, i.e.\ all simultaneous phase symmetries arising from $\Delta(3n^2)$, and Eq.~(\ref{two_triplet_phases}).

This potential has no automatic CP symmetry. Possible CP symmetries are overall $CP_0$ and simultaneous $CP_{23}$, i.e.\ a CP symmetry acting on both triplets with
\begin{equation}
X_{23}^{\varphi \varphi'} = \begin{pmatrix}
X_{23} & 0  \\
0 & X_{23} 
\end{pmatrix},
\label{U23phiphi} \text{ where }
X_{23} = \begin{pmatrix}
1 & 0 & 0 \\
0 & 0 & 1 \\
0 & 1 & 0 
\end{pmatrix}.
\end{equation}
Besides overall $CP_0$ and $CP_{23}$, we will also consider a CP symmetry with 
\be
\label{eq:3n2-Uphiphi'}
X^{\varphi\varphi'} = \begin{pmatrix} X_\varphi & 0 \\0&X_{\varphi'}
\end{pmatrix}
, ~\quad \mathrm{with} ~\quad
X_{\varphi}=\begin{pmatrix} 1&0&0\\
0&\omega&0\\
0&0&\omega^2 \end{pmatrix}
 , \quad
X_{\varphi'}=\begin{pmatrix} 1&0&0\\
0&\omega^2&0\\
0&0&\omega \end{pmatrix}.
\ee
Under $CP_0$, all coefficients of the potential become real such that $\tilde s_3=0$. For $CP_{23}$, $\tilde s_2=0$.
The CP symmetry with $X^{\varphi\varphi'}$, Eq.\ (\ref{eq:3n2-Uphiphi'}), relates parameters of the potential via
\be
\tilde s_3 ~=~ \tilde r_2 ~ i (\omega-\omega^2) \ .
\ee
As $\omega=e^{2\pi i/3}$, we get $\tilde s_3 = -\sqrt{3} \tilde r_2$. 
The CP transformation where the roles of the
explicit matrices in Eq.~\eqref{eq:3n2-Uphiphi'} are exchanged enforces $\tilde s_3 = \sqrt{3}
\tilde r_2$. The effect of the CP symmetries on the VEV orbits will be discussed at the end of this subsection.

Again, arbitrary members of the one-triplet orbits can be combined to pairs which are combined into orbits under the symmetry of the full potential.
Compared to the potential of two triplets of $\Delta(6n^2)$, due to the missing permutation generator in $\Delta(3n^2)$, several orbits split. However, again at this stage it does not make difference to interchange the first and second triplet. For this reason apart from Eqs.~(\ref{2xDelta6n2vevs_start}--\ref{2xDelta6n2vevs_end}), only one new combination survives:
\begin{equation}
 (e^{i\eta},0,0),(e^{i\eta'},0,e^{i\zeta'})\rightarrow (1,0,0),(1,0,1).
\end{equation}
Furthermore, the orbits that contained continuous degrees of freedom now need to be minimized for the more complicated potential invariant under $\Delta(3n^2)$. This is still analytically possible and one obtains

\begin{align}
 &(1,1,0),(1,e^{i\zeta'},0) \text{ and }\\ &(1,1,0),(1,e^{i\zeta'},1)
 \end{align}
with 
\begin{equation}
\zeta'=\arctan(\tilde r_2/\tilde s_3)
\label{eq:zetaarctan}
\end{equation}
a function of $\tilde s_3$ and $\tilde r_2$ in contrast to the situation with a $\Delta(6n^2)$ symmetry, where $\zeta'=0,\pi$, depending on the value of $\tilde r_2$. The orbit with two phases again results in 
\begin{align}
(1,1,1),(1,1,1),\\
(1,1,1),(1,\omega,\omega^2),\\
(1,1,1),(1,\omega^2,\omega),
\end{align}
depending on the values of $\tilde r_2$ and $\tilde s_3$.

To summarize all possible pairs before applying CP symmetries:

\begin{align}
 (1,0,0),(1,0,0)\label{2xDelta3n2vevs_summary}\\
 (1,0,0),(0,1,0)\\
 (1,0,0),(1,1,0)\\
 (1,0,0),(1,0,1)\\
 (1,0,0),(0,1,1)\\
 (1,0,0),(1,1,1)\\
 (1,1,0),(0,1,1)\\
 (1,1,0),(1,e^{i\zeta'},0)\\
(1,1,0),(1,e^{i\zeta'},1)\\
(1,1,1),(1,1,1)\\
(1,1,1),(1,\omega,\omega^2)\\
(1,1,1),(1,\omega^2,\omega)
\end{align}

Next, consider the effects of the various CP symmetries on the orbits. 
For $CP_0$, nothing happens to the the orbits with real representatives. $\tilde s_3=0$, so $\zeta'$ becomes $\pm \pi/2$ and thus $\zeta'=\pm i$. Furthermore, the orbits $(1,1,1),(1,\omega,\omega^2)$ and $(1,1,1),(1,\omega^2,\omega)$ merge.
For $CP_{23}$, $(1,0,0),(1,0,1)$ and $(1,0,0),(1,1,0)$ merge. $\zeta'$ remains free and nothing happens to the orbits with $\omega$s. 
Lastly, for the CP symmetry with $X^{\varphi\varphi'}$, $\tilde s_3=\sqrt{3}\tilde r_2$, thus $\zeta'=\pm \pi/6$. Additionally, $(1,1,1),(1,\omega,\omega^2)$ and $(1,1,1),(1,1,1)$ merge. For the CP symmetry similar to that with $X^{\varphi\varphi'}$ but with block matrices interchanged, $(1,1,1),(1,\omega^2,\omega)$ merges with $(1,1,1),(1,1,1)$.

\subsubsection{$\bf S_4$ \label{sec:S4PPVEVs}}

The $S_4$ potential for one triplet was 
\begin{equation}
\begin{array}{rl}
 V_{S_4} (\varphi) = V_0 (\varphi) 
+b \left[ \left(\varphi_1 \varphi_1 \varphi^{*3} \varphi^{*3}
  +\text{cycl.}\right) + \text{h.c.} \right],
 \end{array}
\end{equation}
with real $b$.
For two triplets we use Eq.~(\ref{eq:potV2}), and write
\bea
V_{S_4} (\varphi,\varphi') &=&
V_0 (\varphi) + V'_0 (\varphi') + V_2 (\varphi,\varphi') +  \label{V24PP}\\[2mm]
&&+~
b \left[\left(
\varphi_1 \varphi_1 \varphi^{*3} \varphi^{*3} + 
\text{cycl.}\right)+\text{h.c.}\right]
 + b' \left[\left(
\varphi'_1 \varphi'_1 \varphi'^{*3} \varphi'^{*3} + 
\text{cycl.} \right)+ \text{h.c.}\right]\notag\\[2mm]
& &+ ~\tilde b \left[ \left(
\varphi_1 \varphi'_1 \varphi^{*3} \varphi'^{*3} + 
\text{cycl.}\right) + \text{h.c.}\right] . \notag
\eea

For $SU(2)$ doublets, we use Eq.~(\ref{eq:potV2H}) and write
\begin{align}
V_{S_4} (H,H') =~& V_0(H) + V_0'(H')+ V_2(H,H') \label{V24HH}\\
&+\sum_{\alpha,\beta} b\left[  \left(
h_{1\alpha} h_{1\beta} h^{\ast3\alpha} h^{\ast3\beta} + 
\text{cycl.} \right)+ \text{h.c.}\right]\notag\\
&+\sum_{\alpha,\beta}  b' \left[\left(
h'_{1 \alpha} h'_{1 \beta} h'^{\ast3\alpha} h'^{\ast3\beta} + 
\text{cycl.}\right) + \text{h.c.} \right]
\notag\\
&+\sum_{\alpha,\beta} \tilde{b}_1\left[ \left(h_{1\alpha}h^{\ast3
    \alpha}h'_{1\beta}h'^{\ast3\beta}+ \text{cycl.}\right)+\text{h.c.}\right]\notag\\
&+\sum_{\alpha,\beta} \tilde{b}_2\left[\left(h_{1\alpha}h'^{\ast 3\alpha}h'_{1\beta}h^{\ast3\beta}+\text{cycl.}\right)+\text{h.c.} \right] . \notag
\end{align}
All $S_4$-invariant potentials listed here, even the two triplet cases, conserve CP explicitly: the parameters, including $b,b',\tilde{b},\tilde{b}_1, \tilde{b}_2$ are all real, so one of the CP symmetries is $CP_0$.

For one triplet of $S_4$, the full symmetries of an even potential are generated by Eq.~(\ref{1xS4_symmetries}),
and notably, this potential has an automatic CP symmetry. The orbits are (and again plus permutations)
\begin{equation}
 \{\begin{pmatrix}\pm e^{i\alpha}\\0\\0\end{pmatrix}\},
 \{\begin{pmatrix}(-1)^k e^{i\alpha}\\(-1)^l e^{i\alpha}\\(-1)^{k+l} e^{i\alpha}\end{pmatrix}\},
 \{\begin{pmatrix}(-1)^k e^{i\alpha}\\\omega(-1)^l e^{i\alpha}\\\omega^2(-1)^{k+l} e^{i\alpha}\end{pmatrix}\},
 \{\begin{pmatrix}-(-1)^k e^{i\alpha}\\\omega(-1)^l e^{i\alpha}\\\omega^2(-1)^{k+l} e^{i\alpha}\end{pmatrix}\},
 \{\begin{pmatrix}0\\\pm e^{i \alpha}\\\pm i e^{i\alpha}\end{pmatrix}\},
\end{equation}
where the last set stands for separate orbits for each combination of signs.
These combine to the following VEVs pairs, where the sign choices correspond to separate orbits (again only orbit representatives)
\begin{align}
(1,0,0),(1,0,0)\\
(1,0,0),(0,1,0)\\
(1,0,0),(1,0,i)\\
(1,0,0),(0,1,i)\\
(1,0,0),(1,1,1)\\
(1,0,0),(1,\omega^2,\omega)\\
(1,0,i),(1,0,\pm i)\\
(1,0,i),(1,i,0)\\
(1,0,i),(1,1,1)\\
(1,0,i),(1,\omega^2,\pm\omega)\\
(1,1,1),(1,1,\pm1)\\
(1,1,1),(1,\pm\omega^2,\omega)\\
(1,\omega^2,\omega),(1,\omega^2,\pm\omega)\\
(1,\omega^2,\omega),(1,-\omega,-\omega^2)\\
(1,\omega^2,\omega),(1,\omega,\omega^2)
\end{align}

\subsubsection{$\bf A_4$ \label{sec:A4PPVEVs}}

In the case of two $A_4$ triplets distinguished by additional symmetries so
that the total symmetry is $A_4\times U(1) \times U(1)'$, the potential
includes a total of seven independent mixed quartic invariants of the form
$\varphi\,\varphi' \, \varphi^* \, \varphi'^*$. 

The $A_4$ symmetric renormalisable
potential takes the following explicit form, with $V_0$ as defined in Eq.~(\ref{eq:potV0}), and using Eq.(\ref{eq:potV1}) we write
\bea
\label{V12PP}
V_{A_4} (\varphi,\varphi') &=&
V_0 (\varphi) + V'_0 (\varphi') + V_1 (\varphi,\varphi') +  \\[2mm]
&&+ \left[
c \left(
\varphi_1 \varphi_1 \varphi^{*3} \varphi^{*3} + 
\text{cycl.} \right) +\text{h.c.}\right]
 + \left[ c' \left(
\varphi'_1 \varphi'_1 \varphi'^{*3} \varphi'^{*3} + 
\text{cycl.} \right)+\text{h.c.}\right]\notag\\[2mm]
 &&+ \left[\tilde c \left(
\varphi_1 \varphi'_1 \varphi^{*3} \varphi'^{*3} + 
\text{cycl.} \right) + \text{h.c.}
\right] .\notag
\eea
The potential is explicitly CP violating \cite{Varzielas:2016zjc} but it is interesting to consider the CP
symmetry where one imposes $X_{23}$ of Eq.~(\ref{U23}) on both triplets, i.e.\ the block matrix $X_{23}^{\varphi \varphi'}$ in Eq.~(\ref{U23phiphi}).
This CP symmetry constrains the potential such that $\tilde s_2=0$, which
forces all explicit CPIs to vanish as expected from the presence of a CP symmetry \cite{Varzielas:2016zjc}.

Furthermore, applying instead the trivial CP symmetry $CP_0$ forces $\tilde
s_3=0$ and all complex parameters ($c,c',\tilde{c}$) to be real. 

For the $SU(2)_L$ version we use Eq.~(\ref{eq:potV1H}) and, for the remainder of the potential $A_4$ potential with two triplets, only the invariant with coefficient $\tilde{c}$ from Eq.~\eqref{V12PP} needs to be doubled:
\begin{equation}\sum_{\alpha,\beta}
\left[ \tilde{c}_1(h_{1\alpha}h^{\ast3 \alpha}h'_{1\beta}h'^{\ast3\beta}+\text{cycl.})+\tilde{c}_2(h_{1\alpha}h'^{\ast 3\alpha}h'_{1\beta}h^{\ast3\beta}+\text{cycl.}) + \text{h.c.}\right].
\end{equation}
We therefore write
\begin{align}
\label{V12HH}
V_{A_4} (H,H') &= V_0(H) + V_0'(H')+ V_1(H,H')\\
&+\sum_{\alpha,\beta}\left[ c \left(
h_{1\alpha} h_{1\beta} h^{\ast3\alpha} h^{\ast3\beta} + 
\text{cycl.} \right)
+ c' \left(
h'_{1 \alpha} h'_{1 \beta} h'^{\ast3\alpha} h'^{\ast3\beta} + 
\text{cycl.} \right) + \text{h.c.} \right]
\notag\\[2mm]
&+\sum_{\alpha,\beta}\left[ \tilde{c}_1(h_{1\alpha}h^{\ast3 \alpha}h'_{1\beta}h'^{\ast3\beta}+ \text{cycl.})+\tilde{c}_2(h_{1\alpha}h'^{\ast 3\alpha}h'_{1\beta}h^{\ast3\beta}+\text{cycl.})+\text{h.c.} \right] \notag.
\end{align}
We note that due to $SU(2)_L$ not allowing cubic invariants of $H$ and/or
$H'$, it is sufficient to use a $Z_3$ symmetry to distinguish the $A_4$
triplets.\footnote{The potential invariant under a
  $Z_2$~\cite{Varzielas:2015joa} would additionally allow for invariants of the form 
$h_{i\alpha}h'^{\ast i \alpha}h_{j \beta}h'^{\ast j   \beta}$ and 
$h_{i\alpha}h'^{\ast i \beta}h_{j \beta}h'^{\ast j   \alpha}$ 
where the conjugated fields are both related to $H'$. } 

It is possible to impose a CP symmetry with 
\bea
X_{23}^{HH'} = \begin{pmatrix}
X_{23} & 0  \\
0 & X_{23} 
\end{pmatrix} \otimes \delta^{\alpha}_{\beta}\ ,
\label{U23HH}
\eea
which, similarly to previous examples, restricts the coefficients in the potential, namely
\begin{equation}
\tilde{s}_{21}=\tilde{s}_{22}=0\ ,
\end{equation}
Imposing, alternatively, the canonical CP symmetry $CP_0$ leads to $\tilde
s_{31}=\tilde s_{32}=0$ as well as $c,c',\tilde{c}_1,\tilde{c}_2\in \mathbb R$.

The full symmetries of an even potential acting on one triplet are generated by Eq.~(\ref{1xA4_symmetries}),
\begin{equation}
 \begin{pmatrix}0&1&0\\0&0&1\\1&0&0\end{pmatrix},
 \begin{pmatrix}e^{i\alpha}&0&0\\0&e^{i\alpha}&0\\0&0&e^{i\alpha}\end{pmatrix},
 \begin{pmatrix}-1&0&0\\0&1&0\\0&0&-1\end{pmatrix},
 \begin{pmatrix}1&0&0\\0&-1&0\\0&0&-1\end{pmatrix}
\end{equation}
and the orbits of one-triplet VEVs are, again with permutations, where now only cyclic permutations are allowed and no longer, as for $S_4$, all possible permutations, and in addition $\beta$ is an arbitrary phase,
\begin{equation}
 \{\begin{pmatrix}\pm e^{i\alpha}\\0\\0\end{pmatrix}\},
 \{\begin{pmatrix}(-1)^k e^{i\alpha}\\(-1)^l e^{i\alpha}\\(-1)^{k+l} e^{i\alpha}\end{pmatrix}\},
 \{\begin{pmatrix}(-1)^k e^{i\alpha}\\\omega(-1)^l e^{i\alpha}\\\omega^2(-1)^{k+l} e^{i\alpha}\end{pmatrix}\},
 \{\begin{pmatrix}-(-1)^k e^{i\alpha}\\\omega(-1)^l e^{i\alpha}\\\omega^2(-1)^{k+l} e^{i\alpha}\end{pmatrix}\},
 \{\begin{pmatrix}0\\\pm e^{i \alpha}\\\pm e^{i\alpha+i\beta}\end{pmatrix}\}.
\end{equation}

The alignment classes listed after this paragraph arise in this case. Note that in the following, the phases $\zeta$ and $\zeta'$ are not arbitrary, but are fixed by the one-triplet parts of the two-triplet potential, as by Eq.~(\ref{A4_alpha_condition}). Where two sign choices are given, they correspond to separate orbits. 
\begin{align}
 (1,0,0),(1,0,0)\\
(1,0,0),(0,1,0)\\
(1,0,0),(1,e^{i \zeta'},0)\\
(1,0,0),(0,1,e^{i \zeta'})\\
(1,0,0),(e^{i \zeta'},0,1)\\
(1,0,0),(1,1,1)\\
(1,0,0),(1,\omega,\omega^2)\\
(1,e^{i \zeta},0),(1,\pm e^{i \zeta'},0)\\
(1,e^{i \zeta},0),(0,1,e^{i \zeta'})\\
(1,e^{i \zeta},0),(e^{i \zeta'},0,1)\\
(1,e^{i \zeta},0),(1,\pm1,1)\\
(1,e^{i \zeta},0),(1,\pm\omega,\omega^2)\\
(1,1,1),(1,1,\pm1)\\
(1,1,1),(1,\omega,\pm\omega^2)\\
(1,1,1),(1,\omega,\omega^2)\\
(1,\omega,\omega^2),(1,\omega,\pm\omega^2)\\
(1,\omega,\omega^2),(1,\omega,\omega^2)
\end{align}

Minima $(1,0,0),(e^{i \zeta'},1,0)$ correspond to a rephasing of the triplets and lie in the same orbit as $(1,0,0),(1,e^{i \zeta'},0)$. Similarly, 
$(1,0,0),(0,e^{i \zeta'},1)$ and
$(1,0,0),(1,0,e^{i \zeta'})$ are already included in the orbits above.

Under trivial CP, orbits with phases merge with their complex conjugates. In particular, as $c,c'=0$, $\zeta,\zeta'=0$.

\subsubsection{$\bf \Delta(54)$ \label{sec:D54PPVEVs}}

Next, consider potentials of two triplets of $\Delta(54)$.
Using Eq.~\ref{eq:potV2} we write
\begin{align}
 V_{\Delta(54)} (\varphi,\varphi')=&V_0(\varphi)+V'_0(\varphi')+V_2(\varphi,\varphi') \label{V54PP}\\&+\left[d \left(
\varphi_1 \varphi_1 \varphi^{*2} \varphi^{*3} + 
\text{cycl.} \right) + \text{h.c.}\right] + \left[d' \left(
\varphi'_1 \varphi'_1 \varphi'^{*2} \varphi'^{*3} + 
\text{cycl.} \right) +\text{h.c.}\right]
 \notag\\&+\tilde d_1\left[   \left(
\varphi_1 \varphi'_1 \varphi^{*2} \varphi'^{*3} + 
\text{cycl.} \right) +\left(
\varphi_1 \varphi'_1 \varphi^{*3} \varphi'^{*2} + 
\text{cycl.} \right) \right]+\text{h.c.}. \notag
\end{align}
Finally, setting
\begin{equation}
  \tilde{d}_{21}=\tilde{d}_{11} \text{ , } \quad 
\tilde{d}_{22}=\tilde{d}_{12}\text{ , }\quad
\tilde{s}_{21}=\tilde{s}_{22}=\tilde{s}_{31}=\tilde{s}_{32}=0,
\end{equation}
we obtain
\begin{align} 
 V_{\Delta(54)} (H,H') &=V_0(H) + V_0'(H')+ V_2(H,H') \label{V54HH}\\&+ 
  \sum_{\alpha,\beta}\left[ d \left(
h_{1\alpha} h_{1\beta} h^{\ast 2 \alpha} h^{\ast 3 \beta}+
\text{cycl.} \right) + 
d' \left(
h'_{1\alpha} h'_{1\beta} h'^{\ast 2 \alpha} h'^{\ast 3 \beta} + 
\text{cycl.} \right)
+ \text{h.c.}\right] \notag
 \\&+  \sum_{\alpha,\beta} \left[
 \tilde{d}_{11}(h_{1\alpha}h^{\ast 2 \alpha}h'_{1\beta}h'^{\ast 3 \beta} +
\text{cycl.})+\tilde{d}_{12}(h_{1\alpha}h'^{\ast 3 \alpha}h'_{1\beta}h^{\ast 2 \beta} +
\text{cycl.}) + \text{h.c.} \right]  \notag
 \\&+  \sum_{\alpha,\beta} \left[  \tilde{d}_{11}(h_{1\alpha}h^{\ast 3 \alpha}h'_{1\beta}h'^{\ast 2 \beta}+
\text{cycl.})+\tilde{d}_{12}(h_{1\alpha}h'^{\ast 2 \alpha}h'_{1\beta}h^{\ast 3 \beta}+
\text{cycl.}) + \text{h.c.} \right] \! . ~~ \notag
\end{align}
The full symmetries of an even potential of one triplet of $\Delta(54)$ were generated by Eq.~(\ref{1xD54_symmetries}). Again, the potential of two triplets of $\Delta(54)$ has no automatic CP symmetries.
The potential of two triplets of $\Delta(54)$ could be reached in several ways: the single-triplet parts of the potential each consist of a $\Delta(6n^2)$ part and a $\Delta(54)$ part, while the same is true for the cross-terms. With Eq.~(\ref{eq:cross}) we can write
\begin{equation}
 V_{\Delta(54)}(\varphi,\varphi')=V_{\Delta(54)}(\varphi)+V'_{\Delta(54)}(\varphi')+V_{\Delta(54),c}(\varphi,\varphi'),
\end{equation}
where each part splits into the part that is symmetric under the larger group and the part that is specific to $\Delta(54)$, e.g. for the cross-terms
\begin{align}
 V_{\Delta(54),c}(\varphi,\varphi')=V_{\Delta(6n^2),c}(\varphi,\varphi')+V'_{\Delta(54),c}(\varphi,\varphi').
\end{align}
The orbits of VEVs of a potential of one triplet of $\Delta(54)$ are
\begin{equation}
 \{\begin{pmatrix}\omega^k e^{i\alpha}\\0\\0\end{pmatrix},\text{perm.}\},\{\begin{pmatrix}\omega^k e^{i\alpha}\\\omega^l e^{i\alpha}\\\omega^{2k+2l} e^{i\alpha}\end{pmatrix},\text{perm.}\},\{\begin{pmatrix}\omega^k e^{i\alpha}\\\omega^l e^{i\alpha}\\\omega^{2k+2l+1} e^{i\alpha}\end{pmatrix},\text{perm.}\},\{\begin{pmatrix}\omega^k e^{i\alpha}\\\omega^l e^{i\alpha}\\\omega^{2k+2l+2} e^{i\alpha}\end{pmatrix},\text{perm.}\}.
 \label{delta54_orbits}
\end{equation}
Here, there are no permutations that are not reproduced by some combination of $(k,l)$. 

The full phase symmetries of two triplets are generated by the direct sum of the generators in Eq.~(\ref{1xD54_symmetries}) except for the potential having two separate phase symmetries for each triplet, equivalently to Eq.~(\ref{two_triplet_phases}). Again combining members of single triplet orbits and using the symmetries of the full potential to reduce the orbits yields
\begin{align}
 &(1,0,0),(1,0,0)\\
 &(1,0,0),(0,1,0)\\ 
 &(1,0,0),(1,1,1)\\ 
 &(1,0,0),(1,1,\omega)\\ 
 &(1,0,0),(1,1,\omega^2)\\
 &(1,1,\omega^i),(\omega^{k'-k},\omega^{l'-l},\omega^{2k'+2l'-2k-2l+i'}) 
 \label{2XD54_vevs}
 \end{align}
Here, the last line stands for several orbits distinguished by the value of $i'$ and $i$.
As all of $k,k',l,l',i,i'=0,1,2$, especially pairs like $(1,\omega,1),(1,1,\omega^2)$ occur, where the phase differences appearing between first and second triplet had previously been unphysical. Nevertheless, as these orbits had only been obtained by reducing the orbits of triplets that already had a $\Delta(54)$ symmetry instead of $\Delta(6n^2)$, the above list is probably incomplete, similarly to the $A_4$ reduction of $S_4$ orbits at the end of section  \ref{1xA4_vev_section}.

The effects of imposing $CP_0$ are analysed in section \ref{2xD54_CP0_section}.

\subsubsection{$\bf \Delta(27)$}

Using Eq.(\ref{eq:potV1}), the potential of two triplets of $\Delta(27)$ even in both triplets is
\bea\label{V27PP}
V_{\Delta(27)} (\varphi,\varphi') &\!\!=\!\!&
V_0 (\varphi) + V_0'(\varphi') + V_1 (\varphi, \varphi') 
  \\[2mm]
&& + \left[d \left(
\varphi_1 \varphi_1 \varphi^{*2} \varphi^{*3} + 
\text{cycl.} \right) + \text{h.c.}\right] + \left[d' \left(
\varphi'_1 \varphi'_1 \varphi'^{*2} \varphi'^{*3} + 
\text{cycl.} \right) +\text{h.c.}\right]
\notag\\[2mm]
&& +\left[ \tilde d_1  \left(
\varphi_1 \varphi'_1 \varphi^{*2} \varphi'^{*3} + 
\text{cycl.} \right) +\text{h.c.}\right] 
+ \left[ \tilde d_2  \left(
\varphi_1 \varphi'_1 \varphi^{*3} \varphi'^{*2} + 
\text{cycl.} \right) +\text{h.c.}\right]
\notag 
 \! .
\eea
Only $d$, $d'$, $\tilde{d}_1$ and $\tilde{d}_2$
are generally complex, the other coefficients are real.
We note also that the $\Delta(54)$ potential (Eq.(\ref{V54PP})) is a particular case of this one, obtained by setting
\bea
\tilde{s}_2 = \tilde{s}_3 = 0,~\tilde{d}_1 = \tilde{d}_2 .
\eea
For $SU(2)_L$ doublets, using Eq.~\ref{eq:potV1H}, the potential is:
\bea
 V_{\Delta(27)} (H,H') \!\!&\!\!\!\!=\!\!\!\!& \!\!V_0(H) + V_0'(H')+ V_1(H,H')+ \label{V27HH}\\
& +&\!\!\sum_{\alpha,\beta}\left[ d \left(
h_{1\alpha} h_{1\beta} h^{\ast 2 \alpha} h^{\ast 3 \beta}+
\text{cycl.} \right) + 
d' \left(
h'_{1\alpha} h'_{1\beta} h'^{\ast 2 \alpha} h'^{\ast 3 \beta} + 
\text{cycl.} \right)
+ \text{h.c.}\right]
\notag\\
&+& \!\!\sum_{\alpha,\beta} \left[
 \tilde{d}_{11}(h_{1\alpha}h^{\ast 2 \alpha}h'_{1\beta}h'^{\ast 3 \beta} +
\text{cycl.})+\tilde{d}_{12}(h_{1\alpha}h'^{\ast 3 \alpha}h'_{1\beta}h^{\ast 2 \beta} +
\text{cycl.}) + \text{h.c.} \right] \notag \\
&+&\!\! \sum_{\alpha,\beta} \left[  \tilde{d}_{21}(h_{1\alpha}h^{\ast 3 \alpha}h'_{1\beta}h'^{\ast 2 \beta}+
\text{cycl.})+\tilde{d}_{22}(h_{1\alpha}h'^{\ast 2 \alpha}h'_{1\beta}h^{\ast 3 \beta}+
\text{cycl.}) + \text{h.c.} \right] \notag
 \! .
\eea

The VEVs of one triplet of $\Delta(27)$ are identical to that of one triplet of $\Delta(54)$, the VEV pairs generated in this way are almost identical for two triplets of $\Delta(27)$ to the above ones of two triplets of $\Delta(54)$. The only effect of the missing permutation in $\Delta(27)$ with respect to $\Delta(54)$ is that several orbits split and that in addition to Eq.~(\ref{2XD54_vevs}), the following pairs represent new independent orbits:
\begin{align}
 &(1,0,0),(1,1,\omega)\\ 
 &(1,0,0),(1,\omega,1)\\ 
 &(1,0,0),(1,1,\omega^2)\\
 &(1,0,0),(1,\omega^2,1).
 \end{align}

\section{Spontaneous CP-odd invariants applied \label{sec:SCPIs}}

Recall that for $SU(2)$ doublets we consider only VEVs that preserve $U(1)_{em}$. By an $SU(2)$ gauge transformation one can without loss of generality assume that for each doublet one component is zero. Due to this, the conclusions about presence or absence of CPV are the same as for the respective potentials of $SU(2)$ singlets, and the SCPI expressions are slightly simpler. For this reason, in this Section we refer only to the potentials of $(\varphi)$ or $(\varphi,\varphi')$.

\subsection{Potentials of one triplet \label{sec:onetrip}}

\subsubsection{$\bf \Delta(3n^2)$ and $\bf \Delta(6n^2)$ with $n>3$ \label{sec:D3_6n2}}

For the potential of one triplet of $\Delta(6n^2)$ or $\Delta(3n^2)$, Eq.~(\ref{eq:potV0}), 
the CP-odd invariants $\mathcal J^{(3,2)}$ and $\mathcal  J^{(3,3)}$ vanish independently of the triplet VEV. As the phases of the allowed VEVs are unphysical, one can always find a CP symmetry that is present in the potential and preserved by the particular VEV. There is no possibility for SCPV.

\subsubsection{$\bf A_4$ and $\bf S_4$ \label{sec:S_A4}}
\cleqn

Even for $A_4$ and $S_4$, the one triplet cases Eq.~(\ref{V12P}) and Eq.~(\ref{V24P}) are fairly trivial, as $\mathcal  J^{(3,2)}$ vanishes before introducing any VEV and $\mathcal J^{(3,3)}$ vanishes after introducing each of the VEVs found. In each case, a CP symmetry preserved by the respective VEV has been found - in particular all the VEVs preserve $X_{23}$, except $(0,1,e^{i\alpha})$, which preserves a rephased version of $X_{23}$ (alternatively, one can perform an unphysical global rephasing of the VEV such that it also preserves $X_{23}$).
 
\subsubsection{$\bf \Delta(27)$ and $\bf \Delta(54)$ \label{sec:S_D27}}

For the discrete symmetries $\Delta(27)$ and $\Delta(54)$, the single triplet potentials Eq.~(\ref{V27P}) and Eq.~(\ref{V54P}) are the same. This potential features the first known case of SGCPV.

In \cite{Varzielas:2016zjc}, already the value of $J^{(3,2)}$ was found for $V_{\Delta(27)} (\varphi)=V_{\Delta(54)} (\varphi)$. For completeness, we repeat it here:
\begin{align}
 \nonumber 
  \mathcal J^{(3,2)}=& \frac{1}{4}(d^{\ast 3}-d^3)(|v_1|^4+|v_2|^4+|v_3|^4-2|v_1|^2|v_2|^2-2|v_1|^2|v_3|^2-2|v_2|^2|v_3|^2)
  \\&+[\frac{1}{2}(dd^{\ast 2}-2d^\ast s^2+d^{2}s)(v_2 v_3 v_1^{\ast 2}+v_1
  v_3 v_2^{\ast2}+v_1 v_2 v_3^{\ast2}) - h.c.]
 \ .\label{eq:J27ex2}
\end{align}
When imposing $CP_0$ (trivial CP, which in this case constrains $\text{Arg}(d)=0$) or any of the other 5 CP transformations listed by \cite{Nishi:2013jqa} which constrain $\text{Arg}(d)$ to be $0$, $2 \pi/3$ or $4 \pi/3$, the expression simplifies to 
\begin{equation}
  \mathcal  J^{(3,2)}_{CP_0} = \frac{1}{2}[dd^{\ast 2}-2d^\ast s^2+d^{2}s] \left[v_2 v_3 v_1^{\ast 2}+v_1 v_3 v_2^{\ast2}+v_1 v_2 v_3^{\ast2} \right]-h.c.  \!.\,
   \end{equation}
As described previously~\cite{Branco:1983tn, deMedeirosVarzielas:2011zw, Varzielas:2012nn, Bhattacharyya:2012pi}, the potential admits minima such as the complex VEV $(1,\omega,\omega^2)$, which is however not CP violating when the potential had trivial CP imposed. It can be confirmed easily that the expression above vanishes for this case - it preserves not $CP_{0}$, but the product of $CP_{0}$ with one of the group elements of $\Delta(27)$ (indeed, $(1,\omega,\omega^2)$ is in the same VEV orbit as $(1,1,1)$, which is real and preserves $CP_0$). Instead, the SGCPV VEV $(\omega,1,1)$ can be inserted into the expression and gives (for $CP_0$ making $d=d^{\ast}$):
\begin{equation}
\mathcal{J}_{CP_0}^{(3,2)}[v(\omega,1,1)]= \frac{3}{2}d(d-s)(d+2s)(\omega - \omega^2) v^4
\end{equation}

For the CP transformation $X_3$ forcing $\text{Arg}(d)=2 \pi/3$ (see Eq.~(\ref{X3})), $d$ is complex, the results are that the VEVs $(0,0,1)$ and $(\omega,1,1)$ preserve some subset of the CP symmetries, so again no SCPV occurs. Instead, the real VEV $(1,1,1)$ and in the same orbit, the complex VEV $(1,\omega,\omega^2)$ show SGCPV~\cite{Fallbacher:2015rea}, as indicated by the SCPI giving $\mathcal{J}_1^{(3,2)}\propto (\omega - \omega^2)$.

The other 6 CP transformations that can be applied to the potential \cite{Nishi:2013jqa}, such as $X_4$ (see Eq.~(\ref{X4})), don't constrain the phase of $d$, but rather relate the parameters $d$ and $s$ such that the SCPI simplifies to:
\begin{equation}
  \mathcal{J}_{X_4}^{(3,2)} = \frac{1}{4}(d^{\ast
    3}-d^3)(|v_1|^4+|v_2|^4+|v_3|^4-2|v_1|^2|v_2|^2-2|v_1|^2|v_3|^2-2|v_2|^2|v_3|^2)\ .
   \end{equation}
The SCPI reveals that for these CP symmetries, SCPV is independent of the phases of the VEV. We verified that the known VEVs for the $CP_0$ symmetric potential, including real VEVs such as $(0,0,1)$, $(1,1,1)$ and complex ones such as $(\omega,1,1)$ are still candidate VEVs of the $X_4$ symmetric potential and all violate CP spontaneously, as indicated by the SCPI not vanishing, e.g.:
\begin{equation}
  \mathcal{J}_{X_4}^{(3,2)}[v(1,0,0)] = \frac{1}{4}(d^{\ast 3}-d^3) v^4
   \end{equation} 

\subsection{Potentials of two triplets \label{sec:twotrip}}

 \subsubsection{$\bf \Delta(3n^2)$ and $\bf \Delta(6n^2)$ with $n>3$ \label{sec:S_D3n2}}

The useful SCPI  $\mathcal J^{(3,2)}$ also takes a non-zero expression for the potential of two triplets of $\Delta(3n^2)$, Eq.~(\ref{V3n2PP}) for $SU(2)_L$ singlets (the respective potential for doublets can be found in Eq.~(\ref{V3n2HH})).

\begin{align}
 \mathcal J^{(3,2)}=
&-\frac{1}{16} \tilde{s}_2 [ \tilde{r}_2  (-4 s-4 s'+2 \tilde{s}_1-\tilde{s}_2+3 \tilde{r}_2) - \tilde{s}_3^2 ] W_{CP_0} \nonumber \\
&-\frac{1}{8} i \tilde{s}_3 \left[\tilde{s}_3^2-3 \tilde{r}_2^2\right] W_{CP_{23}} \nonumber \\
&-\frac{1}{16} i \tilde{s}_2 \tilde{s}_3 [-4s - 4 s' + 2 \tilde{s}_1 -\tilde{s}_2] ({v'}_1 v_2 v_1^\ast {v'}_2^\ast + {v'}_2 v_3 v_2^\ast {v'}_3^\ast  + v_1 {v'}_3 {v'}_1^{\ast} v_3^\ast + h.c.)
\end{align}
where we define VEV dependences for convenience:
 \begin{align} 
W_{CP_{23}} \equiv [ ( |{v'}_1|^2 |{v}_2|^2 + |{v'}_2|^2  |v_3|^2 + |v'_3|^2 |v_1|^2 ) - ( |v_1|^2 |{v'}_2|^2 + |{v}_2|^2 |v'_3|^2 + |v_3|^2 |{v'}_1|^2) ] 
\label{WCP23}
\end{align}
and
\begin{align}
W_{CP_0} \equiv [ ( v_1 {v'}_1^\ast {v'}_2 v_2^\ast + v_2 {v'}_2^\ast {v'}_3  v_3^\ast + v_3 {v'}_3^\ast {v'}_1 v_1^\ast) - h.c.]
\label{WCP0}
\end{align}
which is similar to the remaining VEV dependence apart from the relative minus sign with respect to the hermitian conjugate.

In contrast, for $\Delta(6n^2)$ with two triplets, shown in Eq.~(\ref{V6n2PP}), $\mathcal J_1^{(3,2)}$ vanishes independently of the VEVs, and $\mathcal J_1^{(3,3)}$ takes a non-vanishing expression in general which vanishes when inserting the VEV candidates we found. In each case, all the VEV candidates for the two triplets preserve a CP symmetry (some combination of $CP_0$ or $CP_{23}$), that is automatically present in the general $V_{\Delta(6n^2)}(\varphi, \varphi')$ potential - therefore this potential does not admit SCPV.

\subsubsection{$\bf \Delta(3n^2)$, $n>3$ with $CP_0$}
\label{2xD3n2_CP0_sec}

For $V_{\Delta(3n^2)}$, $n>3$, the non-zero $\mathcal{J}_1^{(3,2)}$ expression becomes meaningful as a measure of SCPV when imposing a CP symmetry on the potential. If we choose $CP_{0}$, then $\tilde{s}_3=0$ and the SCPI becomes:
\begin{align}
 \mathcal{J}_{CP_0}^{(3,2)}=
 -\frac{1}{16} \tilde{r}_2 \tilde{s}_2 [-4 s -4 s' +2 \tilde{s}_1-\tilde{s}_2+3 \tilde{r}_2] W_{CP_0}
\label{eq:J3n2_CP0}
\end{align}
One can note from
\begin{align}
W_{CP_0} \equiv [ ( v_1 {v'}_1^\ast {v'}_2 v_2^\ast + v_2 {v'}_2^\ast {v'}_3  v_3^\ast + v_3 {v'}_3^\ast {v'}_1 v_1^\ast) - h.c.]
\end{align}
that the expression vanishes if the VEVs of the two triplets do not have matching entries (such as $[v(1,0,0),v'(0,1,1)]$). This suggests that those types of VEVs will not show SCPV in this case, regardless of complex phases, confirming that only the relative phase across the same component of the two triplets is physical for this potential. In contrast, a non-zero result is sufficient to show that the VEV pair does SCPV, and the three classes of VEVs that can SCPV in this case are those that in general depend at least on one phase, such as
\begin{align}
\mathcal{J}_{CP_0}^{(3,2)}[v(1,1,0),v'(1,e^{i \zeta'},0)] = \frac{1}{8} i \tilde{r}_2 \tilde{s}_2  (3 \tilde{r}_2-4 s+2 \tilde{s}_1-\tilde{s}_2-4 s') \sin (\zeta') v^2 {v'}^2\\
\mathcal{J}_{CP_0}^{(3,2)}[v(1,1,0),v'(1,e^{i \zeta'},1)] = \frac{1}{8} i \tilde{r}_2 \tilde{s}_2  (3 \tilde{r}_2-4 s+2 \tilde{s}_1-\tilde{s}_2-4 s') \sin (\zeta') v^2 {v'}^2
\end{align}
In this notation, the square brackets denote the pair of VEV considered (including the $v$, $v'$ normalisation), in order to more easily track which VEVs are being plugged in. Note that the ordering of the VEVs can affect the results, e.g.
\be
\mathcal{J}_{CP_0}^{(3,2)}[v(1,1,0),v'(1,e^{i \zeta'},1)] = - \mathcal{J}_{CP_0}^{(3,2)}[v(1,e^{i \zeta'},1),v'(1,1,0)] \,.
\ee
We note also that for $\tilde{s}_3=0$, $\zeta'=\pi/2$ and therefore the VEV of the second triplet becomes $(1,i,0)$ and $(1,i,1)$ respectively (cf.\ Eq.(\ref{eq:zetaarctan})). As the CP violating phases are fixed to geometric values, they are calculable and these cases are considered SGCPV.
The third class that is also very interesting is
\begin{align}
\mathcal{J}_{CP_0}^{(3,2)}[v(1,1,1),v'(1,\omega,\omega^2)] = -\frac{3}{16} \tilde{r}_2 \tilde{s}_2  (3 \tilde{r}_2-4 s+2 \tilde{s}_1-\tilde{s}_2-4 s') (\omega - \omega^2) v^2 {v'}^2
\end{align}
as the phases of the VEVs take special values $\omega$ and $\omega^2$

When the VEV pair representatives take the special values of phases that minimize the potential in certain regions of parameter space, namely one of the triplets aligns in the directions with one entry being $i$, or in the $(1,\omega,\omega^2)$ direction, the SCPI reveals cases with SGCPV.
This is interesting as these are the first reported cases of SGCPV in potentials with 6 fields (arranged here as two triplets), the potential is relatively simple due to the symmetry and the special $\omega$, $\omega^2$ phases appear for a symmetry that is not $\Delta(27)$ or $\Delta(54)$ (the $n=3$ cases).

\subsubsection{$\bf \Delta(3n^2)$, $n>3$ with $CP_{23}$}

When $CP_{23}$ (2-3 swap CP) is imposed on $V_{\Delta(3n^2)} (\varphi,\varphi')$, the respective SCPI $\mathcal{J}_1^{(3,2)}$ expression becomes particularly simple.
$CP_{23}$ for two triplets swaps the 2-3 components of both triplets, i.e.\ $X_{23}^{\varphi \varphi'}$ in Eq.~(\ref{U23phiphi}), which leads to $\tilde{s}_2=0$:
\begin{align}
\mathcal{J}_{CP_{23}}^{(3,2)}=
-\frac{1}{8} i \tilde{s}_3 \left[\tilde{s}_3^2-3 \tilde{r}_2^2\right] W_{CP_{23}}
\label{eq:J3n2_CP23}
\end{align}
where the VEV dependence
 \begin{align} 
W_{CP_{23}} \equiv [ ( |{v'}_1|^2 |{v}_2|^2 + |{v'}_2|^2  |v_3|^2 + |v'_3|^2 |v_1|^2 ) - ( |v_1|^2 |{v'}_2|^2 + |{v}_2|^2 |v'_3|^2 + |v_3|^2 |{v'}_1|^2) ] 
\end{align}
reveals SPCV occurs (or not) regardless of the phases of the VEVs. The three representatives that give non-zero results are proportional to $v^2 {v'}^2$
\begin{align}
\mathcal{J}_{CP_{23}}^{(3,2)}[v(1,0,0),v'(0,1,0)] & \neq 0 \\
\mathcal{J}_{CP_{23}}^{(3,2)}[v(1,0,0),v'(1,0,1)] & \neq 0 \\
\mathcal{J}_{CP_{23}}^{(3,2)}[v(1,1,0),v'(0,1,1)] & \neq 0
\end{align}
featuring three cases with SGCPV. This may appear peculiar given all the phases are zero, but they are calculable and therefore these cases fall under the definition of SGCPV - the (vanishing) phases of the VEVs are stable under small variations of the potential parameters, and there is SCPV.

\begin{figure}[h]
\begin{center}
\includegraphics[scale=0.22]{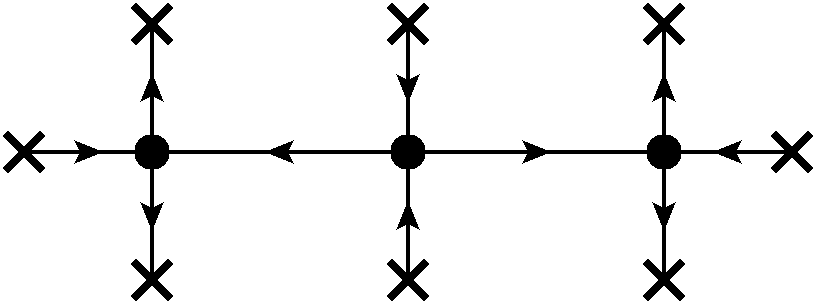}
\end{center}
\caption{The diagram corresponding to the invariant Eq.~(\ref{ZZZvvv17}). }
\label{ZZZvvv17diagram}
\end{figure}

For the type of VEV where the phase $\zeta'$ appears with a physical non-trivial value, CP is spontaneously violated, even though both invariants we are using vanish:
\begin{align}
\mathcal{J}_{CP_{23}}^{(3,2)}[v(1,1,0),v'(1,e^{i\zeta'},0)] = \mathcal{J}_{CP_{23}}^{(3,3)}[v(1,1,0),v'(1,e^{i\zeta'},0)] & = 0 \\
\mathcal{J}_{CP_{23}}^{(3,2)}[v(1,1,0),v'(1,e^{i\zeta'},1)] = \mathcal{J}_{CP_{23}}^{(3,3)}[v(1,1,0),v'(1,e^{i\zeta'},1)] & = 0
\end{align}
As a systematic search for further SCPIs is beyond the scope of this work, instead we have excluded the possibility of $X$ matrices that are simultaneously a symmetry of the potential with general $\tilde s_3\neq0$ and are preserved by the VEVs. The first condition fails for CP symmetries that transform the triplets differently, whereas the second cannot be fulfilled if both triplets transform identically under the CP symmetry. In fact, we were able to find a SCPI which indeed gives a non-zero result for $v(1,1,0),v'(1,e^{i\zeta'},1)$, but as this is the only instance in this paper where it is important, we haven't mentioned it previously. The relevant SCPI is:
\begin{equation}
 J_{CP_{23}}^{(3,4)}=Z^{a_1a_2}_{a_5a_7}Z^{a_3a_4}_{a_6a_8}Z^{a_5a_6}_{a_9a_{10}}v_{a_1}v_{a_2}v_{a_3}v_{a_4}v^{\ast  a_7}v^{\ast  a_8}v^{\ast  a_9}v^{\ast  a_10}
 \label{ZZZvvv17}
\end{equation}
with diagram in Figure \ref{ZZZvvv17diagram}
and we have
\begin{align}
\mathcal{J}_{CP_{23}}^{(3,4)}[v(1,1,0),v'(1,e^{i\zeta'},1)] \propto (2 \sin \zeta' - \sin 2 \zeta').
\end{align}

\subsubsection{$\bf \Delta(3n^2)$, $n>3$ another CP}
\label{2xD3n2_otherCP_section}

In \cite{Varzielas:2016zjc}, explicit CP violation was studied and another class of CP symmetry was found that made the potential CP conserving by relating the parameters $\tilde{s}_3$ and $\tilde{r}_2$ in 3 similar ways. Of these, we denote as \ (cf.\ Eq.(\ref{eq:3n2-Uphiphi'})):
\be
X_{\varphi \varphi'}= \begin{pmatrix} X_{\varphi} & 0 \\0&X_{\varphi'}
\end{pmatrix}
, ~\quad \mathrm{with} ~\quad
X_{\varphi}=\begin{pmatrix} 1&0&0\\
0&\omega&0\\
0&0&\omega^2 \end{pmatrix}
 , \quad
X_{\varphi'}=\begin{pmatrix} 1&0&0\\
0&\omega^2&0\\
0&0&\omega \end{pmatrix},
\ee
this CP symmetry relates the parameters as 
\be
\tilde s_3 ~=~ \tilde r_2 ~ i (\omega-\omega^2),
\ee
i.e.\ $\tilde s_3 = -\sqrt{3} \tilde r_2$.
When this is inserted into the SCPI (considering that $2 \omega = i ( \sqrt{3} + i)$):
\begin{align}
\mathcal{J}_{X_{\varphi \varphi'}}^{(3,2)} =  \frac{1}{8} \tilde{r}_2 \tilde{s}_2 [ 4 s+4 s'-2 \tilde{s}_1+\tilde{s}_2] [ \omega ( v_1 {v'}_1^\ast {v'}_2 v_2^\ast + v_2 {v'}_2^\ast {v'}_3  v_3^\ast + v_3 {v'}_3^\ast {v'}_1 v_1^\ast) - h.c.],
\end{align}
the SCPI becomes similar to the $CP_0$ case but there are relevant factors of $\omega$ and $\omega^2$ which change the VEV dependence crucially (note this is not $\omega W_{CP_0}$, so it is not just an overall multiplicative factor). Indeed, the cases with VEVs featuring powers of $\omega$ that were SGCPV  for $CP_0$ are CP conserving for this case and instead, the pair of VEVs $(1,1,1),(1,1,1)$ is SGCPV. Further, the classes of VEVs with $\zeta'$, which gets fixed to $\zeta' = - \pi/6$ due to the CP symmetry, also gives a non-zero value for the SCPI and are therefore more cases with SGCPV.

\subsubsection{$\bf A_4$ and $\bf S_4$}

The general expression of $\mathcal J^{(3,2)}$ for the $V_{A_4} (\varphi,\varphi')$ potential with two triplets (listed in Eq.~(\ref{V12PP})) is non-zero. We don't show it as it is rather long, and strictly speaking it carries no physical meaning as the potential is explicitly CPV. The expression simplifies sufficiently for relevant cases, when specific CP symmetries are imposed, as shown below as we test each of the twenty two (22) VEV pairs from Section \ref{sec:A4PPVEVs}.

In contrast, $V_{S_4} (\varphi,\varphi')$ (seen in Eq.~(\ref{V24PP})) is a particular case of $V_{A_4} (\varphi,\varphi')$ that automatically preserves both $CP_0$ and $CP_{23}$. The general expression of $\mathcal J^{(3,2)}$ for $V_{S_4} (\varphi,\varphi')$ vanishes, although this does not happen with the SCPI $\mathcal J^{(3,3)}$ for $V_{S_4} (\varphi,\varphi')$. When testing the twenty (20) VEV pairs from Section \ref{sec:S4PPVEVs} we found the SCPI $\mathcal J^{(3,3)}$ always vanishes, which is understood as the VEVs either preserve $CP_0$ (the real VEVs) or $CP_{23}$. In this potential there is no SCPV.

Given this, of the many SCPIs we calculated for the discrete groups $A_4$ and $S_4$, it is only interesting to look in more detail to $V_{A_4} (\varphi,\varphi')$.
   
\subsubsection{$\bf A_4$ with $CP_0$}

When $CP_0$ (trivial CP) is imposed on $V_{A_4} (\varphi,\varphi')$, the respective SCPI $\mathcal J^{(3,2)}$ becomes:
\begin{align}
  \mathcal J_{CP_0}^{(3,2)}=C_{A_4} W_{CP_0}
\label{A4CP0}
\end{align}
with the coefficient dependence
\bea
C_{A_4} = \frac{1}{16} \tilde{s}_2 (\tilde{r}_2 (-3 \tilde{r}_2+4 s-2 \tilde{s}_1+\tilde{s}_2+4 s')-4 \tilde{c} (c+c'))
\eea
and the same VEV dependence, $W_{CP_{0}}$ in Eq.~(\ref{WCP0}), that appeared in the SCPI in Eq.~(\ref{eq:J3n2_CP0}), of the analogous $\Delta(3n^2)$ case invariant under $CP_{0}$. This is not completely unexpected, as the SCPI should vanish regardless of the discrete symmetry being $A_4$ or $\Delta(3n^2)$ with $n>3$, for VEVs that preserve $CP_{0}$ (note though that for the $\Delta(27)$ case, the functional dependence is different).

An analysis of the expression reveals that for this SCPI, only relative phases between the same component of the two triplets appear. This was the case for the $\Delta(3n^2)$ potential, but for $A_4$ there are actually some additional physical phases, which the more complicated SCPI $\mathcal{J}^{(3,3)}$ is sensitive to.

Using $\mathcal{J}^{(3,2)}$ we can confirm that the following VEV pairs SCPV, being proportional to $v^2{v'}^2$:
\begin{align}
\mathcal{J}_{CP_0}^{(3,2)}[v(1,e^{i \zeta},0),v'(1,\pm e^{i \zeta'},0)] & \neq 0 \\
\mathcal{J}_{CP_0}^{(3,2)}[v(1,e^{i \zeta},0),v'(1,\pm 1,1)] & \neq 0 \\
\mathcal{J}_{CP_0}^{(3,2)}[v(1,e^{i \zeta},0),v'(1,\omega,\pm \omega^2)] & \neq 0
\end{align}
and specifically
\begin{align}
\mathcal{J}_{CP_0}^{(3,2)}[v(1,1,1),v'(1,\omega,\omega^2)] &= 3 C_{A_4} (\omega - \omega^2) v^2 {v'}^2\\
\mathcal{J}_{CP_0}^{(3,2)}[v(1,1,1),v'(1,\omega,-\omega^2)] &= C_{A_4} (\omega - \omega^2) v^2 {v'}^2.
\end{align}
The most interesting cases are arguably these latter ones where VEVs only have calculable phases and therefore SGCPV, although one notes they are essentially the same cases obtained for the slightly simpler $\Delta(3n^2)$ potential.

For the cases where the pair of VEVs returns a vanishing expression we confirm if CP is not broken by identifying a CP symmetry preserved by the VEVs (such as $CP_{0}$ itself, for any real VEVs).

\subsubsection{$\bf A_4$ with $CP_{23}$}

When $CP_{23}$ (2-3 swap CP) is imposed on $V_{A_4} (\varphi,\varphi')$, the respective SCPI $\mathcal{J}_1^{(3,2)}$ expression becomes:
\begin{align}
  \mathcal{J}_{CP_{23}}^{(3,2)}=\frac{1}{8} \left[ i \tilde{s}_3 \left(2 \tilde{c}^\ast (c+c')+2 \tilde{c} (c^\ast+c^\ast)+3 \tilde{r}_2^2\right)+2 \tilde{r}_2 (\tilde{c}^\ast (c-c')-c^\ast \tilde{c}+c^\ast \tilde{c})-i \tilde{s}_3^3\right] W_{CP_{23}}.
  \label{A4CP23}
 \end{align}
Note this is the same VEV dependence, $W_{CP_{23}}$ in Eq.~(\ref{WCP23}), that does not depend on the phases of the VEVs and which appeared in the SCPI in Eq.~(\ref{eq:J3n2_CP23}), of the analogous $\Delta(3n^2)$ potential invariant under $CP_{23}$. This is not completely unexpected, as the SCPI should vanish, regardless of the discrete symmetry being $A_4$ or $\Delta(3n^2)$ with $n>3$, for VEVs that preserve $CP_{23}$.

We use this SCPI to check the pairs of VEVs for this potential. We display here only the non-vanishing ones, which are always proportional to $v^2 {v'}^2$, and are:
\begin{align}
\mathcal{J}_{CP_{23}}^{(3,2)}[v(1,0,0),v'(0,1,0)] &\neq 0 \\
\mathcal{J}_{CP_{23}}^{(3,2)}[v(1,0,0),v'(1,e^{i \zeta},0)] & \neq 0 \\
\mathcal{J}_{CP_{23}}^{(3,2)}[v(1,0,0),v'(0,1,e^{i \zeta})] & \neq 0 \\
\mathcal{J}_{CP_{23}}^{(3,2)}[v(1,0,0),v'(e^{i \zeta},0,1)] & \neq 0 \\
\mathcal{J}_{CP_{23}}^{(3,2)}[v(1,e^{i \zeta},0),v'(0,1,e^{i \zeta'})] & \neq 0 \\
\mathcal{J}_{CP_{23}}^{(3,2)}[v(1,e^{i \zeta},0),v'(e^{i \zeta'},0,1)] & \neq 0
\end{align}
The SCPI vanishes for the pair $(1,1,1),(1,\omega,\omega^2)$ regardless of the physical complex phases (this is clear from the SCPI expression, which depends only on the absolute values). It must do so, as the VEV pair preserves $CP_{23}$. In contrast, $[(1,0,0),(0,1,0)]$ is a SGCPV pair of VEVs as the vanishing phases are calculable.

Furthermore, for $A_4$ we have cases entirely analogous to the two VEV pairs that CPV, even though the SCPIs $\mathcal{J}^{(3,2)}$, $\mathcal{J}^{(3,3)}$ vanish:
\begin{align}
\mathcal{J}_{CP_{23}}^{(3,2)}[v(1,1,0),v'(1,e^{i\zeta'},0)] = \mathcal{J}_{CP_{23}}^{(3,3)}[v(1,1,0),v'(1,e^{i\zeta'},0)] & = 0 \\
\mathcal{J}_{CP_{23}}^{(3,2)}[v(1,1,0),v'(1,e^{i\zeta'},1)] = \mathcal{J}_{CP_{23}}^{(3,3)}[v(1,1,0),v'(1,e^{i\zeta'},1)] & = 0
\end{align}
The conclusions are the same as in $\Delta(3n^2)$, as there is no CP symmetry that is both preserved by these VEVs (this condition requires $X$ matrices transforming the two triplets differently) and is a symmetry of the potential with $\tilde s_3\neq0$ (this condition requires $X$ matrices that transform the two triplets equally).

\subsubsection{$\bf \Delta(27)$ and $\bf \Delta(54)$}

For two triplets the potentials $V_{\Delta(27)} (\varphi, \varphi')$ and $V_{\Delta(54)} (\varphi, \varphi')$ are different, as seen in Eq.~(\ref{V27PP}) and Eq.~(\ref{V54PP}) respectively. Nonetheless, the VEV candidates are the same, and the SCPI results are similar. We found several new VEV pairs, many of which are cases with SGCPV.

It turns out the SCPI expressions are slightly more complicated in the $\Delta(27)$ case, but they vanish whenever the corresponding expression in $\Delta(54)$ vanishes, and they are proportional to $(\omega - \omega^2) = i \sqrt{3}$ when the corresponding SCPI expression in $\Delta(54)$ has that dependence. For this reason, we display the SCPI results only for $\Delta(54)$.

\subsubsection{$\bf \Delta(54)$ with $CP_0$}
\label{2xD54_CP0_section}

The last 3 representative VEV pairs from Section \ref{sec:D54PPVEVs} can give non-zero results with SGCPV. In this case we show some cases that make the SCPI vanish, for clarity:

\begin{align}
\mathcal{J}_{CP_0}^{(3,2)}[v(1,1,1),v'(1,\omega,\omega^2)] &= 0\\
\mathcal{J}_{CP_0}^{(3,2)}[v(1,\omega,\omega^2),v'(1,\omega,\omega^2)] &= 0.
\end{align}
The non-vanishing cases include
\begin{align}
\mathcal{J}_{CP_0}^{(3,2)}[v(1,0,0),v' (1,1,\omega)] = - \mathcal{J}_1^{(3,2)}[v(1,0,0),v'(1,1,\omega^2)] = \notag \\
\frac{3}{16} \left(2 \tilde{d}_1^2 ({d'}+s')+4 \tilde{d}_1 \tilde{r}_2 s'+{d'} (8 ({d'}-s') ({d'}+2 s')-\tilde{s}_1 (2 \tilde{r}_2+\tilde{s}_1))\right) (\omega-\omega^2) {v'}^4
\end{align}
and
\begin{align}
\mathcal{J}_{CP_0}^{(3,2)}[v(1,1,\omega),v'(1,\omega,\omega^2)] = - \mathcal{J}_1^{(3,2)}[v(1,1,\omega^2),v'(1,\omega,\omega^2)] = \notag \\
\frac{3}{16} \left(8 d^3+8 d^2 s+d \left(2 \tilde{d}_1^2-\tilde{s}_1 (2 \tilde{r}_2+\tilde{s}_1)-16 s^2\right)+2 \tilde{d}_1 s (\tilde{d}_1+2 \tilde{r}_2)\right) (\omega-\omega^2) v^4 
\end{align}
Note the dependence is either on $v^4$ or ${v'}^4$ exclusively in these cases (and not on $v^2 {v'}^2$). We recall for $\Delta(27)$ and $\Delta(54)$, SCPV is already possible in the single triplet potentials. This contrasts with the $A_4$ and $\Delta(3n^2)$ cases, where only the two triplet potentials admit SCPV - with only two triplets enabling SCPV, the non-vanishing SCPIs for those cases depend always on $v^2 {v'}^2$.
In $\Delta(54)$ we also found VEV pairs for which the SCPI has the $v^2 {v'}^2$ dependence, one such instance is $(1,1,\omega),(1,1,\omega^2)$ which when plugged into the SCPI has all three dependences, $v^4$, ${v'}^4$ and $v^2 {v'}^2$.

\subsection{Spontaneous Geometrical CP Violation}

We have found several new cases of SGCPV. Through these new cases we hope to advance the understanding of what the conditions are for it to occur. We divide our cases mainly in VEVs of one or two triplets where the imposed CP symmetry is $CP_0$ and cases where the imposed CP symmetry is $CP_{23}$.

In the latter case, where $CP_{23}$ is imposed, we found three cases with SGCPV for $\Delta(3n^2)$ with $n>3$:
\begin{align}
\mathcal{J}_{CP_{23}}^{(3,2)}[v(1,0,0),v'(0,1,0)] & \neq 0 \\
\mathcal{J}_{CP_{23}}^{(3,2)}[v(1,0,0),v'(1,0,1)] & \neq 0 \\
\mathcal{J}_{CP_{23}}^{(3,2)}[v(1,1,0),v'(0,1,1)] & \neq 0
\end{align}
and one similar case for $A_4$:
\begin{align}
\mathcal{J}_{CP_{23}}^{(3,2)}[v(1,0,0),v'(0,1,0)] & \neq 0.
\end{align}
It occurs always in situations with two triplets where the VEVs have the calculable phase of zero (i.e.\ the VEVs are real), in pairs that simultaneously violate $CP_{23}$ and the cyclic permutations of $CP_{23}$ (which we may refer to as $CP_{31}$ and $CP_{12}$) that are also CP symmetries of the potential when $CP_{23}$ is imposed, due to the flavour symmetries we are considering. Indeed, it is understandable why there is no SGCPV of this type for the respective one triplet potentials, as a real single triplet VEV cannot violate all 3 of those CP symmetries. We note though that the real VEV pairs that SGCPV in these cases would have preserved $CP_0$, which is not however a symmetry of the potentials (otherwise there wouldn't be SCPV).

In the cases where $CP_0$ is imposed, which include also the original SGCPV cases with one triplet of $\Delta(27)$ (or $\Delta(54)$), we note also that the VEVs with calculable phases that have SGCPV would have preserved some other ``calculable'' CP symmetry \footnote{By extension of the definition of calculable phases \cite{Branco:1983tn}, we refer to a calculable CP symmetry as one that is independent of the parameters of the potential.} that is not a symmetry of the potential. In the one triplet case, take:
\begin{equation}
\mathcal{J}_{CP_0}^{(3,2)}[v(\omega,1,1)]= \frac{3}{2}d(d-s)(d+2s)(\omega - \omega^2) v^4
\end{equation}
this VEV would have preserved a variant of the $X_2$, $X_3$ CP transformations:
\begin{equation}
 X=\begin{pmatrix}\omega&0&0\\0&1&0\\0&0&1\end{pmatrix}
 \end{equation} 
The SGCPV VEVs in $\Delta(3n^2)$:
\begin{align}
\mathcal{J}_{CP_0}^{(3,2)}[v(1,1,0),v'(1,i,0)] = \frac{1}{8} i \tilde{r}_2 \tilde{s}_2  (3 \tilde{r}_2-4 s+2 \tilde{s}_1-\tilde{s}_2-4 s') v^2 {v'}^2\\
\mathcal{J}_{CP_0}^{(3,2)}[v(1,1,0),v'(1,i,1)] = \frac{1}{8} i \tilde{r}_2 \tilde{s}_2  (3 \tilde{r}_2-4 s+2 \tilde{s}_1-\tilde{s}_2-4 s') v^2 {v'}^2
\end{align}
would have preserved a CP with diagonal matrix $X =diag(1,1,1,1,-1,1)$ and
\begin{align}
\mathcal{J}_{CP_0}^{(3,2)}[v(1,1,1),v'(1,\omega,\omega^2)] = -\frac{3}{16} \tilde{r}_2 \tilde{s}_2  (3 \tilde{r}_2-4 s+2 \tilde{s}_1-\tilde{s}_2-4 s') (\omega - \omega^2) v^2 {v'}^2
\end{align}
would have preserved $CP_{23}$. $CP_{23}$ would also be preserved by the first of the SGCPV pair of VEVs in $A_4$:
\begin{align}
\mathcal{J}_{CP_0}^{(3,2)}[v(1,1,1),v'(1,\omega,\omega^2)] &= 3 C_{A_4} (\omega - \omega^2) v^2 {v'}^2\\
\mathcal{J}_{CP_0}^{(3,2)}[v(1,1,1),v'(1,\omega,-\omega^2)] &= C_{A_4} (\omega - \omega^2) v^2 {v'}^2
\end{align}
The second pair, due to the minus sign, would have preserved the CP symmetry with a block matrix with the regular $X_{23}$ for the first triplet and for the second triplet:
\be
X_{-23} = \begin{pmatrix} 1&0&0\\
0&0&-1\\
0&-1&0 \end{pmatrix}.
\ee
The new SGCPV VEVs we find in $\Delta(54)$ (and $\Delta(27)$) are essentially two triplet versions of the one triplet case:
\begin{align}
\mathcal{J}_{CP_0}^{(3,2)}[v(1,0,0),v' (1,1,\omega)] \neq 0\\
\mathcal{J}_{CP_0}^{(3,2)}[v(1,1,\omega),v'(1,\omega,\omega^2)] \neq 0\\
\mathcal{J}_{CP_0}^{(3,2)}[v(1,1,\omega),v' (1,1,\omega^2)] \neq 0
\end{align}
and for each case one can find $X$ matrices of the calculable CP symmetries that would have been preserved.
With one triplet the other case with SGCPV that we presented is
\begin{equation}
  \mathcal{J}_{X_4}^{(3,2)}[v(1,0,0)] = \frac{1}{4}(d^{\ast 3}-d^3) v^4
   \end{equation} 
and this VEV would have preserved $CP_{0}$.
Finally, the other CP symmetry we studied for two triplets of $\Delta(3n^2)$ with $n>3$ led to the following cases of SGCPV:
\begin{align}
\mathcal{J}_{CP_X}^{(3,2)}[v(1,1,0),v'(1,e^{-i \pi/6},0)] \neq 0\\
\mathcal{J}_{CP_X}^{(3,2)}[v(1,1,0),v'(1,e^{-i \pi/6},1)] \neq 0\\
\mathcal{J}_{CP_X}^{(3,2)}[v(1,1,1),v'(1,1,1)] \neq 0.
\end{align}
Of these, the last VEV pair would clearly have preserved $CP_0$, and the other VEVs preserve a CP with diagonal matrix $X =diag(1,1,1,1,e^{i \pi/3},1)$ featuring only calculable phases ($0$ and $\pi/3$).

All of the cases found that show SGCPV have calculable phases in VEVs that preserve a different calculable CP symmetry (e.g. real VEVs that violate $CP_{23}$ but would have preserved $CP_{0}$). We conjecture that this is a requirement for SGCPV to occur.
It is important to stress that the CP symmetries that these VEVs would have preserved are themselves calculable, not depending on the parameters of the potential.
For cases with SCPV that is not geometric one can find CP symmetries that would have been preserved by the respective VEVs, but those CP symmetries will also not be calculable, depending on the parameters of the potential, which run with energy scale.

This usefulness of this conjecture is illustrated by our examples. In cases where it is possible to identify distinct calculable CP symmetries that can be imposed to constrain the parameters of a specific potential, that potential can then give rise to SGCPV, depending of course on the VEVs allowed by the respective minimisation conditions.

\section{Summary of results \label{sec:sum}}

In this Section we summarise the results presented throughout Section \ref{sec:SCPIs}, in two tables. Table \ref{ta:1_summary} lists the potentials with one triplet of the discrete symmetries and Table \ref{ta:2_summary} the potentials with two triplets. For both tables, the leftmost column notes what is the imposed discrete symmetry and, in cases where it is also imposed, the CP symmetry. The next two columns show the two SCPIs we calculated and distinguish whether the expression vanishes in general (before plugging in the VEVs) - note that in some of the cases, despite the expression not vanishing in general, it vanishes for all the VEVs.
The fourth column lists if a CP symmetry is present, listing the matrix associated with the CP symmetry of one triplet (for two triplets, this means it is the diagonal $2\times2$ block matrix with the same transformation for both triplets) - this column is relevant because in many cases where there was no CP symmetry imposed, there is nevertheless a CP symmetry present for the potential that is invariant under the imposed discrete flavour symmetry. The last column notes whether there is Spontaneous CP Violation. In this last column we note 2 categories when there is no imposed CP symmetry - the potential is either CP conserving, and all of these cases have no SCPV either\footnote{These results support a conjecture from \cite{Ivanov:2014doa}, proven for specific cases in \cite{Branco:2015bfb}.}, or alternatively the potential is explicitly CP violating and therefore we denote the SCPV column with NA for Not Applicable. The final possibility is when a CP symmetry is imposed, and it turned out that for all the cases analysed with these discrete symmetries, whenever there was the possibility of SCPV, there were some (but not all) VEVs that have SGCPV, and therefore this column is marked as ``S(G)CPV'', because that potential can have regular SCPV or SGCPV depending on the VEVs (we note though that this is probably due to the discrete symmetries we analyse being special, as many potentials, such as the 2HDM with imposed CP symmetry, can have SCPV but don't have SGCPV).

In summary:
\begin{itemize}
 \item We present new results for $V_{\Delta(3n^2)} (H) = V_{\Delta(6n^2)} (H)$.
 \item Of the 1 triplet potentials we studied, only $V_{\Delta(27)} (H) = V_{\Delta(54)} (H)$ has non-trivial CP properties, which  includes cases with Spontaneous Geometrical CP Violation.
 \item All the 2 triplet results we present are new.
 \item Of the 2 triplets potentials we studied, all that can have SCPV, also show explicit CPV.
 \item For the potentials we studied where SCPV can be found, some VEVs exist with SGCPV (this is a peculiarity of the symmetries we considered, as it is not true in general).
 \item We have found several new cases of SGCPV beyond the few known cases up to now, finding also the first cases of SGCPV for 6HDMs.
 \item We formulate a conjecture relating SGCPV with calculable CP symmetries.
\end{itemize}

\begin{table}[t]
\centering
\begin{tabular}{|c||c|c||c|c|}
\hline
 &$J^{(3,2)}$& $J^{(3,3)}$&\text{CP}&SCPV\\ \hline \hline
 ${\bf3}_{\Delta(3n^2)=\Delta(6n^2)}$&0&0& $X_0$,$X_{23}$ & No\\ \hline
  \hline
 ${\bf3}_{A_4}$&0&*& $X_{23}$ & No\\\hline
 ${\bf3}_{S_4}$&0&*& $X_0$, $X_{23}$ & No\\\hline
 \hline
 ${\bf3}_{\Delta(27)=\Delta(54)}$&*&*& None & NA\\\hline
 ${\bf3}_{\Delta(27)=\Delta(54)}$, $X_0$&*&*& $X_{0}$ & S(G)CPV\\\hline
 ${\bf3}_{\Delta(27)=\Delta(54)}$, $X_{4}$&*&*& $X_{4}$ & S(G)CPV\\\hline

\end{tabular}
\caption{Summary of the value of $J^{(3,2)}$, $J^{(3,3)}$, CP symmetry transformations and whether SCPV is possible, for the 1 triplet scalar potentials analysed. The $*$ denotes that a SCPI is non-zero for arbitrary VEVs (it may still vanish for the potential's VEVs), NA stands for Not Applicable (SCPV is not meaningful in cases without a CP symmetry).  \label{ta:1_summary}}
\end{table}

\begin{table}[t]
\centering
\begin{tabular}{|c||c|c||c|c|}
\hline
 &$J^{(3,2)}$& $J^{(3,3)}$&\text{CP}&SCPV\\ \hline \hline

 $2 \times  {\bf3}_{\Delta(3n^2)}$&*&*& None & NA\\ \hline
  $2 \times  {\bf3}_{\Delta(3n^2)}$,$CP_0$&*&*& $X_0$ & S(G)CPV\\ \hline
  $2 \times  {\bf3}_{\Delta(3n^2)}$, $CP_{23}$&*&*& $X_{23}$ & S(G)CPV\\ \hline
  $2 \times  {\bf3}_{\Delta(3n^2)}$, $CP_X$ &*&*& $X'$ & S(G)CPV\\ \hline
 $2 \times  {\bf3}_{\Delta(6n^2)}$&0&*& $X_0$,$X_{23}$ & No\\ \hline
\hline
 $2 \times {\bf3}_{A_4}$&*&*& None & NA\\\hline
  $2 \times {\bf3}_{A_4}$, $CP_0$&*&*& $X_{0}$ & S(G)CPV\\\hline
   $2 \times {\bf3}_{A_4}$, $CP_{23}$&*&*& $X_{23}$ & S(G)CPV\\\hline
 $2 \times {\bf3}_{S_4}$&0&*& $X_0$, $X_{23}$ & No\\\hline
   \hline
 $2 \times  {\bf3}_{\Delta(27)}$&*&*& None & NA\\\hline
   $2 \times  {\bf3}_{\Delta(27)}$, $CP_0$&*&*& $X_0$ & S(G)CPV\\\hline
 $2 \times  {\bf3}_{\Delta(54)}$&*&*& None & NA\\\hline
   $2 \times  {\bf3}_{\Delta(54)}$, $CP_0$&*&*& $X_0$ & S(G)CPV\\\hline
\end{tabular}
\caption{Summary of the value of $J^{(3,2)}$, $J^{(3,3)}$, CP symmetry transformations and whether SCPV is possible, for the 2 triplet scalar potentials analysed. The $*$ denotes the SCPI is non-zero for arbitrary VEVs (it may still vanish for the potential's VEVs), NA stands for Not Applicable (SCPV is not meaningful in cases without a CP symmetry).  \label{ta:2_summary}}
\end{table}

\section{Conclusions \label{sec:con}}

We have analysed 3 and 6 Higgs scalar potentials invariant under discrete symmetries $\Delta(3n^2)$ and $\Delta(6n^2)$ with $n=2$ ($A_4$, $S_4$), $n=3$ ($\Delta(27)$, $\Delta(54)$) and $n>3$. For these potentials, we have presented their VEVs and considered whether they can have Spontaneous CP Violation, and if so, if it can be Spontaneous Geometrical CP Violation.

Concerning the minimisation of such complicated potentials, the strategy we have followed in this paper may be 
summarised as follows. We started with more symmetric cases, which can be minimised analytically, where each of the minima of these potentials is associated with an orbit, which is generated by the representative VEV by the symmetry group. For the less symmetric potentials, we have analysed how these orbits are broken up and which phases can become physical as a result of reducing the original symmetry. We find that for potentials involving one triplet, this method reproduces perfectly the minima found by a previous thorough analysis done by other authors. We then proceeded to construct the orbits of potentials of two triplets, again starting from more symmetric cases, which in this case means the potentials of two triplets without cross-terms. We then used the conjectured minima in the subsequent analysis of CPV.

With respect to Spontaneous CP Violation, using the basis invariant formalism, we presented two Spontaneous CP-odd Invariants and used these to confirm cases where there is Spontaneous CP Violation.
With our methods, we have confirmed several existing results in the literature for the 3 Higgs potentials, 
i.e.\ involving one triplet under the specified symmetries of $A_4$, $S_4$, $\Delta(27)$ and $\Delta(54)$. Beyond that, we present new results for the potentials invariant under $\Delta(3n^2)$ and $\Delta(6n^2)$ with $n>3$ and those with 6 Higgs, i.e.\ involving two triplets under the above symmetries.
Of the potentials considered, for those that were automatically CP invariant we found no VEVs that spontaneously violate CP - all the SCPIs vanish and we found for each VEV at least one preserved CP symmetry. This is in line with what was the case for 3 Higgs \cite{Ivanov:2014doa} and the more general conjecture proven in the particular case of rephasing symmetries in \cite{Branco:2015bfb}.

For potentials that in general have explicit CP violation, however, we considered imposing different CP symmetries and then checking which (if any) VEVs spontaneously violate CP. After revisiting the well-known case for the 3 Higgs potential invariant under both $\Delta(27)$ and trivial CP (where SGCPV was first identified) we also found SGCPV in the 6 Higgs potentials invariant under $A_4$, $\Delta(27)$, $\Delta(54)$ and $\Delta(3n^2)$ with $n>3$. These new cases were found by finding a non-zero SCPI and checking that the VEVs in question have calculable phases. We have proposed the following conjecture:
SGCPV appears when there are different calculable CP symmetries that can be imposed on a potential, and the VEV violates the imposed calculable CP symmetry but would have preserved a different calculable CP symmetry.

Finally we remark that the results in this paper may be applied to physical CP violating processes involving three and six-Higgs doublet potentials
controlled by classes of non-Abelian discrete symmetries. 

\section*{Acknowledgements}
The authors thank Igor Ivanov and Renato Fonseca for useful discussions.

IdMV acknowledges
funding from Funda\c{c}\~{a}o para a Ci\^{e}ncia e a Tecnologia (FCT) through the
contract IF/00816/2015 and partial support from the HARMONIA project under contract
UMO-2015/18/M/ST2/00518 (2016-2019).
This project has received funding from the European Union's Seventh Framework Programme for research, technological development and demonstration under grant agreement no PIEF-GA-2012-327195 SIFT.

The work of CL is
supported by the Deutsche Forschungsgemeinschaft (DFG) within the
Research Unit FOR 1873 ``Quark Flavour Physics and Effective Field
Theories''.

SFK acknowledges the STFC Consolidated Grant ST/L000296/1 and the European Union's Horizon 2020 Research and Innovation programme under Marie Sk\l{}odowska-Curie grant agreements Elusives ITN No.\ 674896 and InvisiblesPlus RISE No.\ 690575. 

Work  supported  by  MINECO  grants  FPA2014-58183-P, Multidark CSD2009-00064, and the PROMETEOII/2014/084 grant from Generalitat Valenciana.

\newpage

\appendix

\section{Technical details from section \ref{sec:CPV}}
\label{technical_details_appendix}

In this appendix, first the argument leading to Eq.~(\ref{VEV_CP_conservation_new}) is given. After that, additionally a related relation between VEVs is derived.
From each allowed CP transformation, conditions on $Y$ and $Z$ follow,
\begin{equation}
 Y^a_b=Y^{b'}_{a'}X_{b'b}X^{\ast a'a}
 \label{Y_CP_trafo_new}
\end{equation}
and
\begin{equation}
 Z^{ab}_{cd}=Z^{c'd'}_{a'b'}X_{c'c}X_{d'd}X^{\ast a'a}X^{\ast b'b}.
\label{Z_CP_trafo_new}
 \end{equation}
When the potential ``acquires'' a VEV, i.e.\ when the parameters of the potential are chosen such that the energy is classically minimized at a value of the field different from zero,
\begin{equation}
\langle\phi\rangle=v\neq0, 
\end{equation}
then first the fields and with them the potential can be expanded around this VEV:
\begin{equation}
 \varphi\mapsto v+\varphi
\end{equation}
and
\begin{equation}
V(\phi)\equiv V(v,\vp)=Y^a_b (v+\vp)_a (v+\vp)^{\ast b}+Z^{ab}_{cd}(v+\vp)_{a}(v+\vp)_{b}(v+\vp)^{\ast c}(v+\vp)^{\ast d}.
\end{equation}
This potential can be reordered by the number of fields $\vp$, i.e.\
\begin{align}
 V(v,\vp)=&Y^a_bv_av^{\ast b}+Y^a_bv_a\vp^{\ast b}+Y^a_b\vp_av^{\ast b}+Y^a_b\vp_a\vp^{\ast b}\nonumber\\
 &+Z^{ab}_{cd}v_av_bv^{\ast c}v^{\ast d}+\ldots+Z^{ab}_{cd}\vp_a\vp_b\vp^{\ast c}\vp^{\ast d}\nonumber\\
 =&Y^a_bv_av^{\ast b}+Z^{ab}_{cd}v_av_bv^{\ast c}v^{\ast d}\nonumber\\
 &+(Y^a_bv^{\ast b}+\ldots+Z^{ab}_{cd}v_bv^{\ast c}v^{\ast d} )\vp_a\nonumber\\
 &+\ldots\nonumber\\
 &+Z^{ab}_{cd}\vp_a\vp_b\vp^{\ast c}\vp^{\ast d}.
\end{align}
The coefficients of this potential are now combinations of $Y$, $Z$, and $v$.
This potential is CP-conserving in the new degrees of freedom, $\vp$, if a CP trafo
\newcommand{\tx}{\ti X}
\begin{equation}
 \vp\mapsto \ti X \vp^\ast
\end{equation}
exists, such that the minimized potential is invariant,
\begin{equation}
 V(v,\vp)=V(v,\ti X \vp^\ast).
\end{equation}
From this, conditions on the coefficients of the expanded potential follow. For the coefficient quartic in $\vp$, this happens to be the original condition, only with $X$ replaced with $\tx$,
\begin{equation}
 Z^{ab}_{cd}\tx_{a}^{a'}\tx_{b}^{b'}\tx^{\ast c}_{c'}\tx^{\ast d}_{d'}={Z^{\ast}}^{a'b'}_{c'd'},
\end{equation}
while for the trilinear coefficient in $\vp$, a new condition arises,
\begin{equation}
 Z^{ab}_{cd}v_a \tx_b^{b'}\tx^{\ast c}_{c'}\tx^{\ast d}_{d'}={Z^\ast}^{eb'}_{c'd'}v_e
 \label{trilinear_invariance}
\end{equation}
together with the CP-conjugate of this condition. (Similarly for lower powers of $\vp$, only with more complicated conditions that involve combinations of $Y$ and $Zvv$.)

Can there be $\tx$ that are not in $\mathcal{X}=\{X\}$? First, as the condition for CP-conservation in the quartic terms in $\phi$ and $\vp$ are identical, and all CP transformations of $\phi$ were assumed to be known, there cannot be any more general CP transformations under which the whole potential is invariant than these known ones. In other words, all candidates for CP transformations of $\vp$, $\tx$ are in $\mathcal{X}=\{X\}$.

Using the invariance of $Z$ on the LHS of Eq.\ (\ref{trilinear_invariance}) leads to 
\begin{equation}
 {Z^\ast}^{a'b'}_{c'd'}{X^\dagger}_{a'}^av_a={Z^\ast}^{a'b'}_{c'd'}v_{a'}
\end{equation}
from which with ${X^\dagger}^a_{a'}v_a=(v_{a'})^\ast$ follows that the potential conserves CP if for at least one of the original $X$ matrices holds that
\begin{equation}
 v_{a}=X_{aa'}v^{\ast a'}.
\end{equation}

\textit{Which conditions follow on VEVs or on relations between VEVs from an existing CP symmetry with unitary matrix $X$ at high energy?} Transform the potential with said CP transformation to obtain
\begin{equation}
 V=Y^a_b X_{aa'}X^{\ast b b'}\phi^{\ast a'} \phi_{b'}+Z^{ab}_{cd}X_{aa'}X_{bb'}X^{\ast cc'}X^{\ast dd'} \phi^{\ast a'}\phi^{\ast b'}\phi_{c'}\phi_{d'}.
\end{equation}
From this potential, minimisation conditions can be derived, and from
\begin{equation}
 \frac{\partial V}{\partial \phi_i}=0
\end{equation}
follows
\begin{equation}
 0=Y^a_i X_{aa'}v^{\ast a'}+2 Z^{ab}_{id}X_{aa'}v^{\ast a'}X_{bb'}v^{\ast b'}X^{\ast dd'}v_{d'}.
\end{equation}
Comparing this with the CP-conjugate of the untransformed minimisation condition, i.e.\ the condition obtained by taking the derivative of $V$ by $\phi^{\ast i}$,
\begin{equation}
 0=Y^a_i v^a+ 2Z^{ab}_{id}v_a v_b v^{\ast d},
 \label{eom1}
\end{equation}
one finds that $v_a$ and $u_a:=X_{aa'}v^{\ast a'}$ fulfill exactly the same sets of equations. This means that the solution sets, $\{v_a\}$ and $\{X_{aa'}v^{\ast a'}\}$ are identical:
\begin{equation}
 \{v_a\}=\{X_{aa'}v^{\ast a'}\}=: \mathcal{V}.
\end{equation}
Note that this does \textit{not} mean that for every $v_a$ holds that $v_a=X_{aa'}v^{a'}$, but instead, as $X$ is unitary, the only thing that can be said is that for symmetric $X$ for every $v_a \in \mathcal{V}$  exactly one $u_a\in \mathcal{V}$ exists, such that 
\begin{equation}
 v_a=X_{aa'}u^{\ast a'}
 \label{X_mapping}
\end{equation}
and longer chains for non-symmetric $X$ matrices.

\end{document}